\documentclass[11pt,prd,onecolumn,showpacs,amsmath,amssymb,aps,floats,floatfix,nofootinbib]{revtex4-1}
\usepackage{hyperref} 
\usepackage{dcolumn}
\usepackage{bm}
\usepackage{amsmath}
\usepackage{amsfonts}
\usepackage{amssymb}
\usepackage{color}
\usepackage{latexsym}
\usepackage{slashed} 
\usepackage{pstricks}
\usepackage{indentfirst}
\usepackage{mathrsfs}
\usepackage{multirow}
\usepackage{epsfig,psfrag}
\usepackage{subfigure}
\usepackage{setspace} 

\graphicspath{{figs/}}

\def\sigmav{\langle \sigma_{\mathrm{ann}} v \rangle}
\def\cm{\mathrm{cm}} 
\def\sec{\mathrm{s}} 
\def\ifb{\mathrm{fb}^{-1}} 
\def\TeV{\mathrm{TeV}}     
\def\GeV{\mathrm{GeV}}     
\def\MeV{\mathrm{MeV}}     
\def\pT{p_\mathrm{T}} 
\def\missET{\slashed E_\mathrm{T}} 
\def\mueff{\mu_\mathrm{eff}}  
\def\tanb{\tan\beta}
\def\m{\mathrm{m}}
\def\mT{m_\mathrm{T}}
\def\mTT{m_\mathrm{T2}}
\def\msfos{m_\mathrm{SFOS}}
\def\br{\mathrm{BR}}
\def\aa{a_1}
\def\ha{h_1}
\def\hb{h_2}
\def\chibc{\tilde{\chi}_{2,3}^0}
\def\chia{\tilde{\chi}_1^0}
\def\chib{\tilde{\chi}_2^0}
\def\chic{\tilde{\chi}_3^0}

\def\chiapm{\tilde{\chi}_1^{\pm}}

\makeatother

\allowdisplaybreaks 

\begin{document}

\setstretch{1.2} 

\title{Searching for Singlino-Higgsino Dark Matter in the NMSSM}
\author{Qian-Fei Xiang$^1$}
\author{Xiao-Jun Bi$^1$}
\author{Peng-Fei Yin$^1$}
\author{Zhao-Huan Yu$^2$}
\affiliation{$^1$Key Laboratory of Particle Astrophysics,
Institute of High Energy Physics, Chinese Academy of Sciences,
Beijing 100049, China}
\affiliation{$^2$ARC Centre of Excellence for Particle Physics at the Terascale,
School of Physics, The~University of Melbourne, Victoria 3010, Australia}

\begin{abstract}
We study a simplified scenario in the next-to-minimal supersymmetric standard model with a split electroweak spectrum, in which only the singlino and higgsinos are light and other superpartners are decoupled.
Serving as a dark matter candidate, a singlino-dominated neutralino $\tilde{\chi}_1^0$ should have either resonant annihilation effects or sizable higgsino components to satisfy the observed relic abundance.
The sensitivities of LHC searches and dark matter detection experiments are investigated. With an integrated luminosity of $30~(300)~\mathrm{fb}^{-1}$, $3l + \missET$ and $2l + \missET$ searches at the 13~(14)~TeV LHC are expected to reach up to $m_{\chia}\sim 150~(230)~\mathrm{GeV}$ and $m_{\tilde{\chi}_2^0,\tilde{\chi}_1^{\pm}}\sim 320~(480)~\mathrm{GeV}$.
Near future dark matter direct and indirect detection experiments are promising to cover the parameter regions where collider searches lose their sensitivities.
\end{abstract}
\maketitle

\section{Introduction}

With the discovery of the Higgs boson at the Large Hadron Collider (LHC)~\cite{Aad:2012tfa,Chatrchyan:2012xdj}, the complete particle content of the Standard Model (SM) has been experimentally confirmed.
However, the large radiative correction to the Higgs mass term leads to the hierarchy problem, which implies that there should be new physics between the electroweak scale and the Planck scale.
In addition, the SM cannot explain the existence of dark matter (DM) in the Universe.
Therefore, new particles as the DM candidate are required in new physics beyond the SM.
Among numerous new physics scenarios, supersymmetry (SUSY) provides an elegant solution to the hierarchy problem by introducing the contributions to the Higgs mass term from superpartners.
Moreover, the lightest supersymmetric particle (LSP) in the $R$-parity conserved SUSY models is absolutely stable and could be an excellent DM candidate.

The SUSY extension with the minimal field content is known as the Minimal Supersymmetric Standard Model (MSSM), which has many attractive features but also faces some challenges.
For instance, the reason why the dimensional parameter $\mu$ in the supersymmetric mass term $\mu\hat{H}_u \hat{H}_d$ is far below the Planck scale is not explained in the MSSM. This is the well known ``$\mu$-problem''~\cite{Kim:1983dt}.
Moreover, the mass of the lighter CP-even neutral higgs is subject to a constraint, $m_h^2 \le m_Z^2 \cos^2 2 \beta$, at the tree level.
Although loop effects can lift the mass up to $\sim 125~\GeV$ to meet with the observed value, it is somewhat fine-tuned~\cite{BasteroGil:2000bw} and put some constraints on the particle spectrum.
For instance, the third generation squarks are required to be light in the SUSY (see e.g.~\cite{Kitano:2006gv,Papucci:2011wy,Cao:2012fz,Kang:2012sy}).

The Next-to-Minimal Supersymmetric Standard Model (NMSSM) solves the $\mu$-problem by adding a singlet chiral superfield $\hat{S}$ to the MSSM (see Refs.~\cite{Maniatis:2009re,Ellwanger:2009dp} for recent reviews).
As a result, $\mu$ is replaced by a dynamical quantity $\mueff = \lambda v_s$ when $S$ develops a VEV $v_s$, which is naturally at the electroweak scale.
Furthermore, the mass of the SM-like Higgs can be easily interpreted due to the enlarged Higgs sector, which contains three CP-even neutral Higgs bosons, two CP-odd neutral Higgs bosons, and two charged Higgs boson.

Since no superpartner has been found, SUSY searches at the LHC have set stringent constraints on the masses of superpartners. In particular, the masses of gluinos and the first two generations of squarks are required to be much higher than 1~TeV~\cite{Aad:2015iea,Aad:2015baa,Khachatryan:2015vra}. The constraints on the masses of neutralinos and charginos are much weaker due to the small electroweak production cross sections.
For instance, in the case of pure wino $\chib$ and $\chiapm$ with pure bino $\chia$, the ATLAS limit $\m_{\chib} \gtrsim 350~\GeV$ are obtained for $\m_{\chia} \lesssim 100~\GeV$, assuming $\br (\chib \to Z \chia) = 100 \%$~\cite{Aad:2014nua}. Exclusion limits from other LHC searches for the electroweak superpartners can be found in Refs.~\cite{Aad:2014vma,Aad:2015eda,Aad:2015jqa,Khachatryan:2014qwa,Khachatryan:2014mma}.
These limits are derived in some simplified scenarios assumed. Thus, in a realistic MSSM they would be changed due to reduced branching ratios and modified kinematics~\cite{Fowlie:2013oua,Gori:2013ala,Han:2013kza,Schwaller:2013baa,Das:2014kwa,Martin:2014qra,Calibbi:2014lga,Han:2014xoa,Han:2014sya,diCortona:2014yua,Chakraborti:2015mra,Nelson:2015jza,Bramante:2015una,Badziak:2015qca,Chakraborti:2015mra,Chakraborty:2015xia,Cao:2015efs,Choudhury:2016lku,vanBeekveld:2016hbo}.
In the NMSSM, neutralinos have additional singlino components $\tilde S$ from the fermionic part of $\hat{S}$. As a result, the interpretation of the LHC SUSY searches, as well as the DM phenomenology, would be affected (see e.g.~\cite{Gunion:2005rw,Belanger:2005kh,Hugonie:2007vd,Cao:2011re,Cheng:2013hna,Cheng:2013fma,Kozaczuk:2013spa,Badziak:2015exr}).

In this work, we focus on the case where the LSP is a singlino-dominated neutralino (see e.g.~\cite{Ellwanger:2013rsa,Kim:2014noa,Dutta:2014hma,Cao:2014efa,Enberg:2015qwa,Potter:2015wsa}). The mass hierarchy among the bino, winos, higgsinos, and singlino is controlled by the diagonal elements of the neutralino mass matrix: $M_1$, $M_2$, $\mueff$, and $2\kappa v_s$, where $\kappa$ comes from the singlet self-interaction term $\frac{1}{3}\kappa \hat{S}^3$.
Since the pure singlino DM would be overproduced in the early Universe due to the limited singlino interactions, the LSP $\chia$ should have some other components to provide an acceptable relic abundance for a standard cosmology. We can define three simplified scenarios for the singlino-dominated LSP: the singlino-bino scenario ($2\kappa v_s < M_1 \ll M_2, \mueff$), the singlino-wino scenario ($2\kappa v_s < M_2 \ll M_1, \mueff$), and the singlino-higgsino scenario ($2\kappa v_s < \mueff\ll M_1, M_2$).
In these scenarios, some particular gauginos or higgsinos with much higher masses decouple from the rest superpartners, leading to specific phenomenological consequences.
Because there is no mixed mass term between the singlino and the bino/wino, the singlino can only mix with the bino/wino via higgsino states.
Therefore, $\chia$ has a very large singlino component in the singlino-bino and singlino-wino scenarios, and cannot easily explain the observed DM relic density.

In order to satisfy the observed relic density, the bino-like LSP is usually required to be lighter than $100~\GeV$~\cite{ArkaniHamed:2006mb}.
In the singlino-bino scenario, $\chia$ could be even lighter, e.g. $m_{\chia} \sim \mathcal{O}(10)~\GeV$~\cite{Gunion:2005rw,Cao:2011re,Cao:2013mqa}.
Although $\chia$ and $\chib$ can be quite light, the production rates of $\chia\chib$ and $\chib\chib$ at the LHC would still be very low compared with SM backgrounds.
If other electroweak superpartners are too heavy, it will be difficult to explore this scenario through electroweak production at the LHC. The singlino-wino scenario is analogous to the bino-wino scenario with similar definition in the MSSM. In this scenario, the correct relic abundance could be achieved when there occurred coannihilation between $\chia$ and $\chiapm/\chib$ in the early Universe, which requires a strong mass degeneracy.
For such a squeezed spectrum, final state leptons from $\chiapm\chib$ production would be soft, and hence a hard initial state radiation jet could be helpful.
For the bino-wino scenario, LHC $3l$ searches are expected to reach $m_{\chia} \sim 220~(320)~\GeV$ for $m_{\chib} - m_{\chia} = 20~(30)~\GeV$ at $\sqrt{s} = 14~\TeV$ with an integrated luminosity of $300~\ifb$~\cite{Gori:2013ala}. These limits could be approximately applied to the singlino-wino scenario.

Below we will only focus on the singlino-higgsino scenario, where the LSP $\chia$ is mainly singlino, while $\chibc$ and $\chiapm$ are mainly higgsinos. Some recent works on this scenario include studies on LHC searches~\cite{Ellwanger:2013rsa,Kim:2014noa} and IceCube indirect searches~\cite{Enberg:2015qwa}. In this work, first we will investigate the viable parameter regions and decay patterns of neutralinos and charginos. Then we will derive current bounds and future prospects of LHC searches, DM direct detection, and DM indirect detection.
In order to have effective $\chia\chia$ annihilation in the early Universe,
the mixture with higgsinos and annihilation through a Higgs/$Z$ boson resonance would be helpful~\cite{Nihei:2002ij,Belanger:2005kh}. Near the resonance regions, it is difficult to probe $\chia$ in direct detection experiments because the effective DM couplings to quarks might drop dramatically.
As higgsino-dominated $\chibc$ and $\chiapm$ are light, LHC searches in the $3l+\missET$ and $2l+\missET$ final states would be sensitive to DM signatures through $\chibc \chiapm$ and $\tilde{\chi}_1^{+} \tilde{\chi}_1^{-}$ pair production processes, respectively. However, if the mass splitting between $\chibc / \chiapm$ and $\chia$ is small, the LHC sensitivity would decrease due to the low reconstruction efficiency of soft leptons. In this case, because $\chia$ has moderate higgsino components, it remains possible to probe DM in direct and indirect detection experiments.

This paper is organized as follows.
In Sec.~\ref{sec:scan} we provide details and results of a parameter scan in the singlino-higgsino scenario and present three typical benchmark points.
Sec.~\ref{sec:LHC} focuses on LHC searches in the $3l+\missET$ and $2l+\missET$ channels.
In Sec.~\ref{sec:exps} we investigate the sensitivity of DM detection experiments.
Sec.~\ref{sec:concl} gives our conclusions and discussions.

\section{Parameter space scan}
\label{sec:scan}

The $Z_3$-invariant NMSSM superpotential is~\cite{Maniatis:2009re,Ellwanger:2009dp}
\begin{equation}
  W=W_\mathrm{MSSM} + \lambda \hat{S} \hat{H_u} \hat{H_d} + \frac{1}{3} \kappa \hat{S}^3,
\end{equation}
where $W_\mathrm{MSSM}$ is the MSSM superpotential, and $\lambda$ and $\kappa$ are dimensionless couplings.
Once $S$ develops a VEV $v_s$, an effective $\mu$-term, $\mueff \hat{H_u} \hat{H_d}$, is generated with $\mueff = \lambda v_s$.
The soft breaking terms in the Higgs sector are given by
\begin{equation}
V_\mathrm{soft} = m_{H_u}^2 |H_u|^2 +  m_{H_d}^2 |H_d|^2 + m_S^2 |S|^2 +\left(\lambda A_\lambda S H_u H_d + \frac{1}{3} \kappa A_\kappa S^3 +\mathrm{h.c.}\right).
\label{eq:potential}
\end{equation}
The minimization of the scalar potential relates the soft parameters $m_{H_u}^2$, $ m_{H_d}^2$, and $ m_S^2$ to $m_Z$, $v_s$, and $\tanb \equiv v_u/v_d$, where $v_u$ and $v_d$ are the VEVs of $H_u$ and $H_d$.
Therefore, the Higgs and higgsino sectors in the NMSSM are determined by 6 parameters:
\begin{equation}
  \lambda,~ \kappa,~ A_\lambda,~ A_\kappa,~ \mueff,~ \tanb.
  \label{eq:six}
\end{equation}

In the gauge basis $\psi_\alpha = (\tilde{B}, \tilde{W}^0, \tilde{H}_d^0, \tilde{H}_u^0, \tilde{S})$,
the neutralino mass term can be expressed as $- \dfrac{1}{2} [\psi_\alpha (M_{\tilde{\chi}^0})_{\alpha\beta} \psi_\beta+\mathrm{h.c.}]$, where the symmetric mass matrix is
\begin{equation}
  M_{\tilde{\chi}^0} =
    \left(
    \begin{array}{ccccc}
      M_1  & 0 & -g_1 v_d/\sqrt{2} & g_1 v_u/\sqrt{2}   & 0 \\
      ~  & M_2 & g_2 v_d/\sqrt{2} & -g_2 v_u/\sqrt{2}   & 0 \\
      ~  & ~ & 0 & -\mueff   & -\lambda v_u \\
      ~  & ~ & ~ & 0   & -\lambda v_d \\
      ~  & ~ & ~ & ~   & 2 \kappa v_s
    \end{array}
    \right).
  \label{eq:M}
\end{equation}
It can be diagonalized by a mixing matrix $N$, and hence the mass eigenstates are related to the gauge eigenstates through
\begin{equation}
  \tilde{\chi}_i^0 = N_{i1}\tilde{B} + N_{i2}\tilde{W}^0+ N_{i3}\tilde{H}_d^0
  + N_{i4}\tilde{H}_u^0 + N_{i5}\tilde{S}.
  \label{eq:chia}
\end{equation}

For the singlino-higgsino scenario, we perform a random scan to identify the NMSSM parameter points that satisfy the observed DM relic density.
In order to reduce the number of free parameters, we fix $M_1$, $M_2$, and $M_3$ to be $2~\TeV $, $ 2~\TeV$, and $5~\TeV$, respectively.
Moreover, all trilinear couplings and soft mass terms for squarks and sleptons are set to be $5~\TeV$.
Thus the bino, winos, gluinos, squarks, and sleptons will be heavy and decouple from the physics we concern.
The remaining free parameters are related to the Higgs and higgsino sectors.
We carry out a random scan within the following ranges:
\begin{eqnarray}
 &&  100~\GeV \le \mueff \le 600~\GeV,~~
    -1~\TeV \le A_\kappa \le 0,~~
    100~\GeV \le A_\lambda \le 10~\TeV,
 \\
 &&  1 \le \tanb \le 50,~~
    0.05 \le \lambda \le 0.7,~~
    0.05 \le \kappa \le 0.7.
  \label{le:para}
\end{eqnarray}
Here we require that the singlino-dominated $\chia$ should satisfy $|N_{15}|^2 > 0.5$. A recent comprehensive study on the allowed NMSSM parameter space can be found in Ref.~\cite{Beskidt:2016egy}.

We employ the package \texttt{NMSSMTools~4.6.0}~\cite{Ellwanger:2004xm,Ellwanger:2005dv,Ellwanger:2006rn} for calculating particle spectra, decay branching ratios, and many other observables.
DM relic density, direct detection, and indirect detection results are computed through the embedded \texttt{micrOMEGAs~3} code~\cite{Belanger:2013oya}.
During the scan, several constraints are imposed as follows.
\begin{description}
\item[DM relic density]
the $\chia$ relic density $\Omega_{\chia} h^2$ is required to be below $0.131$, consistent with the latest Planck measurement~\cite{Planck:2015xua}.
\item[Higgs bounds]
one of the Higgs scalar should be SM-like and its mass should be within the range of $122-128~\GeV$ \footnote{Here we adopt a default setting of NMSSMTools (\emph{Option 8 0}) to calculate the Higgs masses without including the full loop corrections. Once these corrections to Higgs masses are considered (\emph{Option 8 2} for the full one-loop corrections and the two-loop $\mathcal{O}(\alpha_t\alpha_s+\alpha_b\alpha_s)$ corrections), some benchmark points we selected might not satisfy all the Higgs bounds.
}.
Its couplings to other SM particles should be consistent with the results derived from a global fit to the measurements of the Higgs partial decay widths within $3\sigma$ derivations (see e.g.~\cite{Bernon:2014vta}).
\item[LEP bounds]
direct SUSY searches at the LEP have set bounds on superpartners.
Here we impose two relevant bounds.
One is that the lighter chargino should satisfy $m_{\chiapm} > 103.5~\GeV$, which is determined by the LEP collision energy.
The other one is that the $Z$ invisible width should satisfy $\Gamma_Z^\mathrm{inv} < 2~\MeV$ at $95\%$ CL~\cite{ALEPH:2005ab}.
When the decay channel into $\chia\chia$ opens, this width may exceed the experimental value.
\item[Muon $g-2$]
a light Higgs would significantly affect the muon anomalous magnetic moment $a_\mu=(g_\mu-2)/2$, whose most accurate measurement comes from the E821 experiment~\cite{Bennett:2004pv}.
Here we require the NMSSM contribution within the $3 \sigma$ derivation, i.e., $-5.62 \times 10^{-11} < a_\mu^\mathrm{NMSSM} < 5.54 \times 10^{-9}$.
\item[B physics bounds]
there are flavor constraints from $B$ meson rare decays, such as $B_s\to \mu^+\mu^-$, $B^+ \to \tau^{+} \nu$, and $B_s\to X_s\gamma$.
We use the recent experimental results at $95\%$ CL: $1.7 \times 10^{-9} < \br(B_s \to \mu^{+} \mu^{-}) < 4.5 \times 10^{-9}$~\cite{Amhis:2014hma}, $0.85 \times 10^{-4} < \br(B^+ \to \tau^{+} \nu) < 2.89 \times 10^{-4}$~\cite{Lees:2012ju}, and $2.99 \times 10^{-4} < \br(B_s\to X_s\gamma) < 3.87 \times 10^{-4}$~\cite{Amhis:2014hma}.
\end{description}

\begin{figure}[!htbp]
\centering
\subfigure[~$\Omega_{\chia} h^2$ vs. $m_{\chia}$~\label{fig:omega:a}]{
\includegraphics[width=0.45\textwidth]{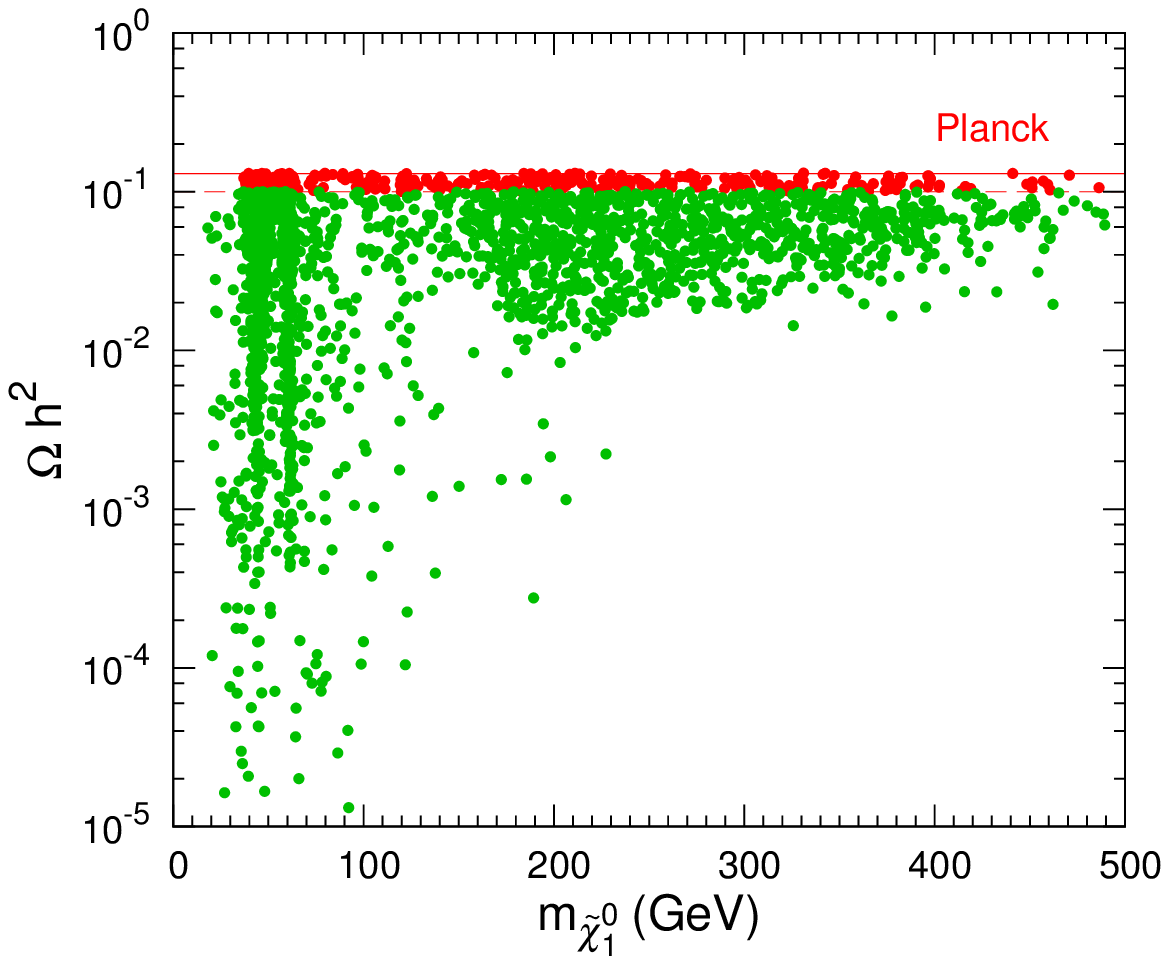}}
\subfigure[~$|N_{15}|^2$ vs. $m_{\chia}$~\label{fig:omega:b}]{
\includegraphics[width=0.45\textwidth]{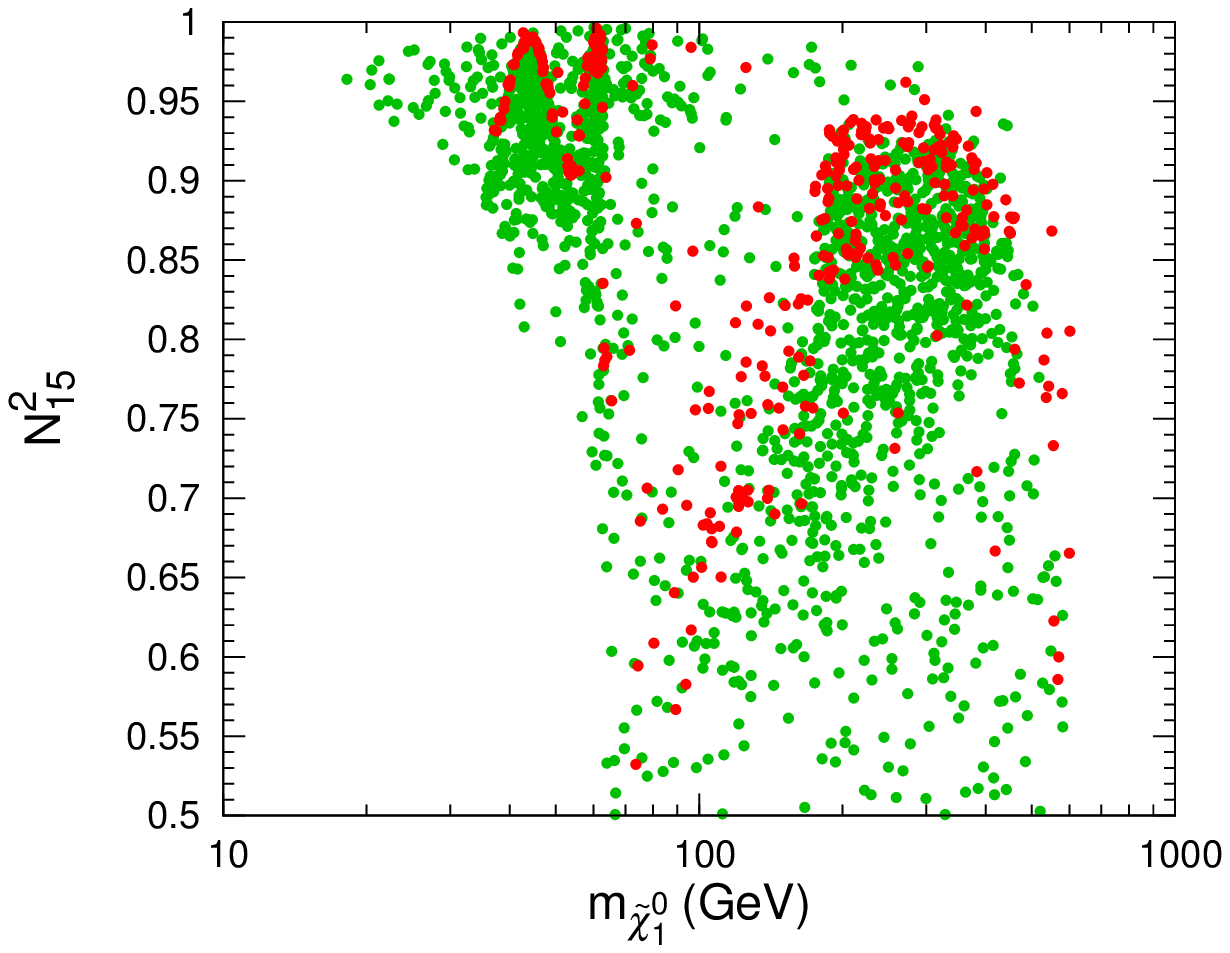}}
\caption{DM relic density $\Omega_{\chia} h^2$ (a) and singlino component $|N_{15}|^2$ (b) versus the LSP mass $m_{\chia}$.
All points satisfy $\Omega_{\chia} h^2 < 0.131$,
while the red points also satisfy $\Omega_{\chia} h^2 > 0.107$.}
\label{fig:omega}
\end{figure}

Now we analyze the properties of the parameter points survived from the above constraints.
Fig.~\ref{fig:omega:a} shows the calculated $\chia$ relic density for a standard cosmology.
The red points can saturate the observed relic abundance ($0.107<\Omega_{\chia} h^2 <1.131$), while the green points predict a lower abundance ($\Omega_{\chia} h^2 <0.107$), which may be compensated by other production mechanisms, e.g., nonthermal production~\cite{Lin:2000qq,Fujii:2002kr,Kane:2015qea} and cosmological enhancement due to the quintessential effect~\cite{Salati:2002md,Rosati:2003yw,Profumo:2003hq}.
Another possibility is that $\chia$ may just constitute a fraction of the whole DM~\cite{Zurek:2008qg}.
Fig.~\ref{fig:omega:b} shows $|N_{15}|^2$, the squared singlino component of $\chia$.
This confirms that the survived points truly correspond to singlino-dominated $\chia$.

Two bunches of points gather around $m_{\chia} \sim 45~\GeV$ and $\sim 60~\GeV$, corresponding to resonance enhancements of the $Z$ boson and the SM-like Higgs boson for $\chia\chia$ annihilation, respectively.
There are also some scattered points yielding a very low relic density, due to resonance enhancements of other Higgs scalars, whose masses are undetermined.
In addition, most of the points with $m_{\chia} \gtrsim 70~\GeV$ do not have resonance effects. In this case, $\chia$ has larger higgsino components and a smaller singlino component.

\begin{figure}[!htbp]
\centering
\subfigure[~$\chib$ decay branching ratios~\label{fig:branch:a1}, $h_1$ is SM-like Higgs boson]{
\includegraphics[width=0.45\textwidth]{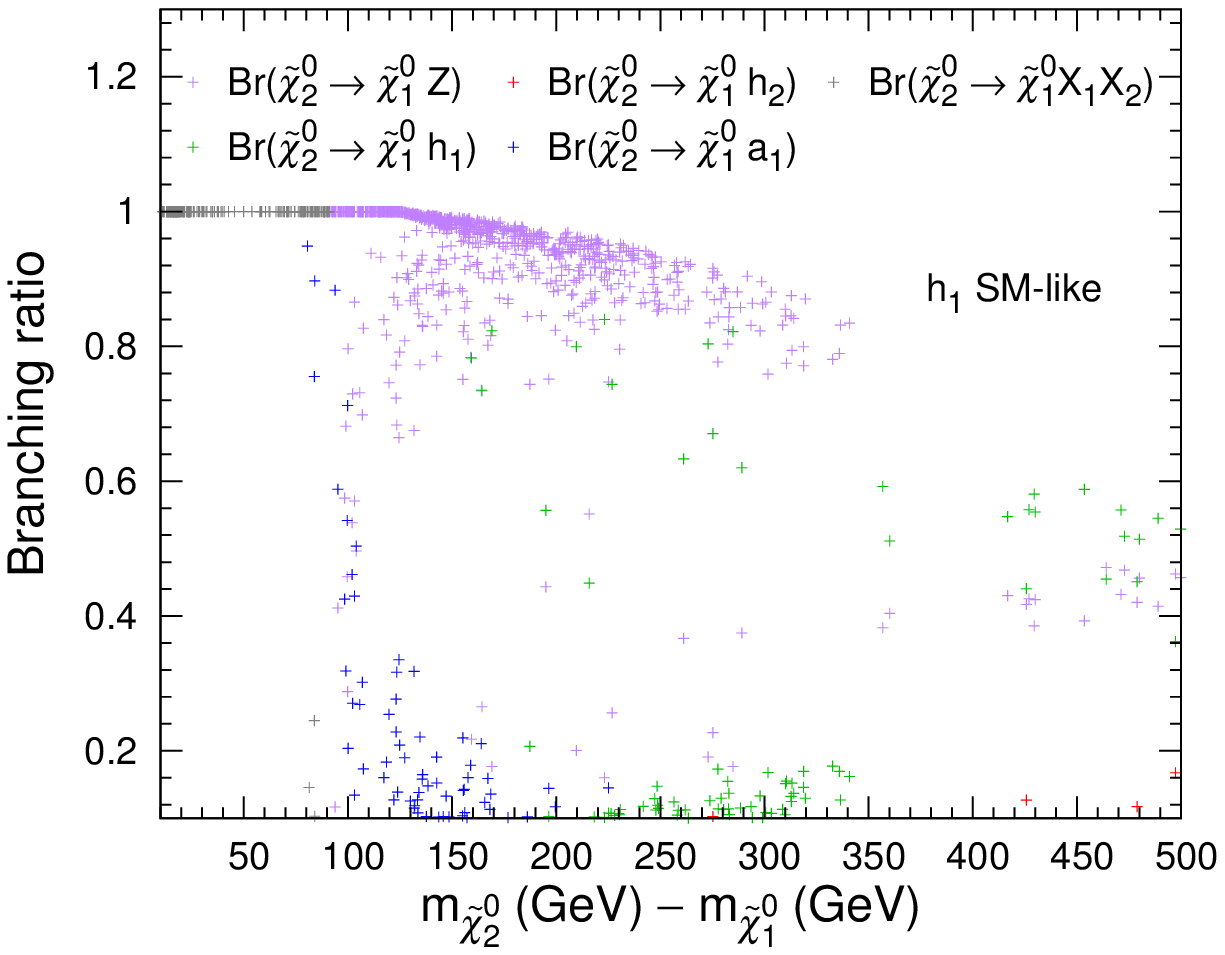}}
\subfigure[~$\chib$ decay branching ratios~\label{fig:branch:a2}, $h_2$ is SM-like Higgs boson]{
\includegraphics[width=0.45\textwidth]{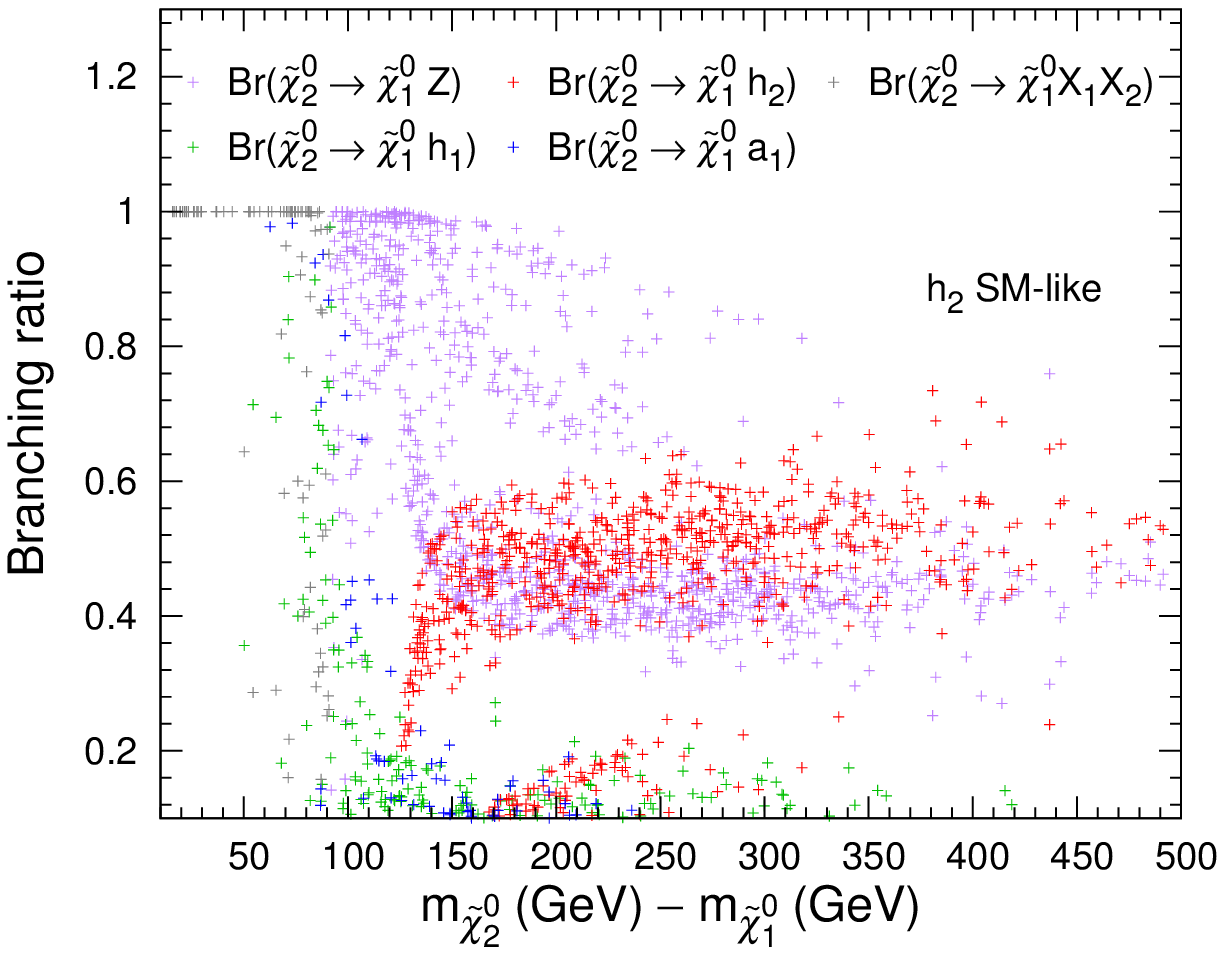}}
\subfigure[~$\chic$ decay branching ratios~\label{fig:branch:b1}, $h_1$ is SM-like Higgs boson]{
\includegraphics[width=0.45\textwidth]{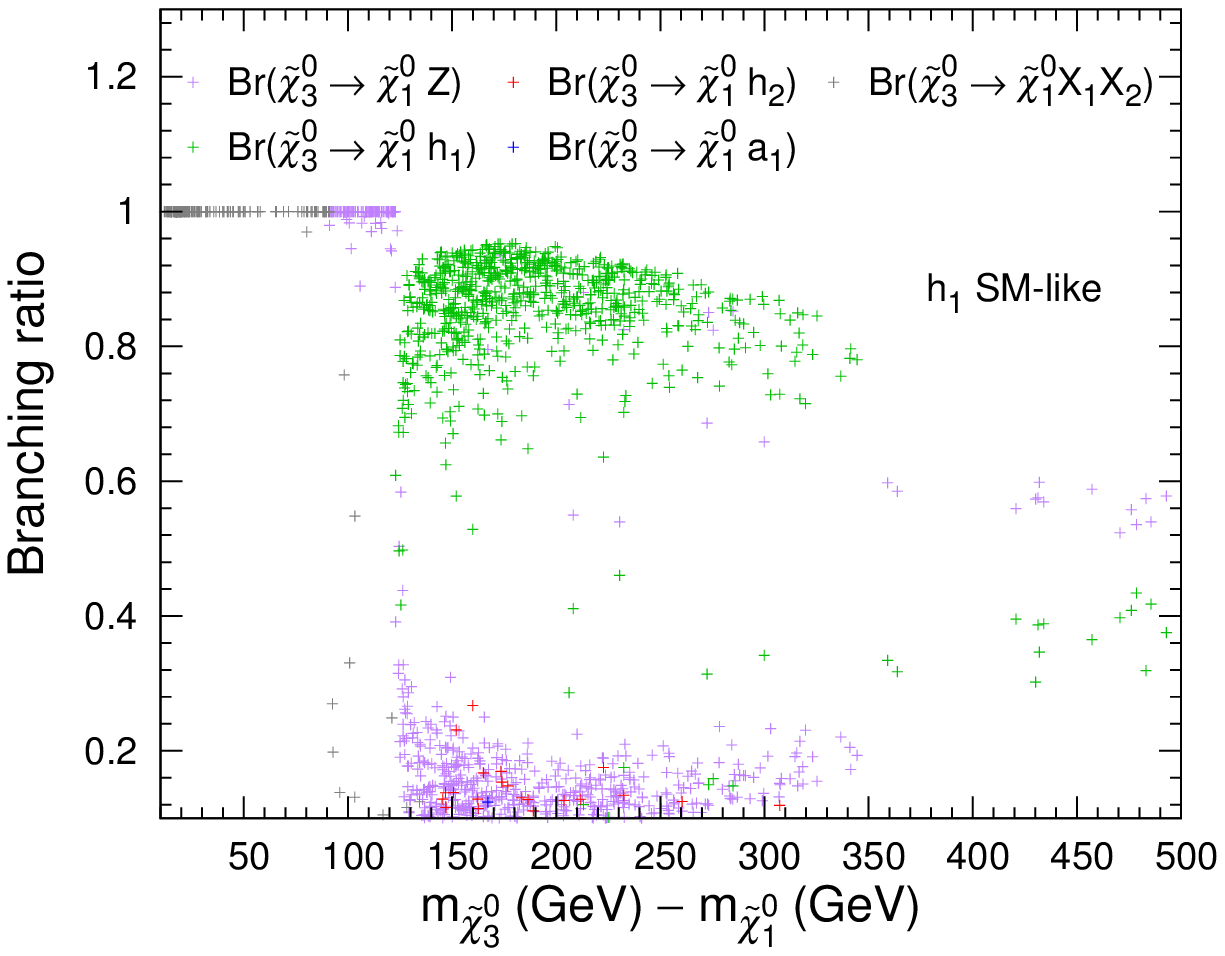}}
\subfigure[~$\chic$ decay branching ratios~\label{fig:branch:b2}, $h_2$ is SM-like Higgs boson]{
\includegraphics[width=0.45\textwidth]{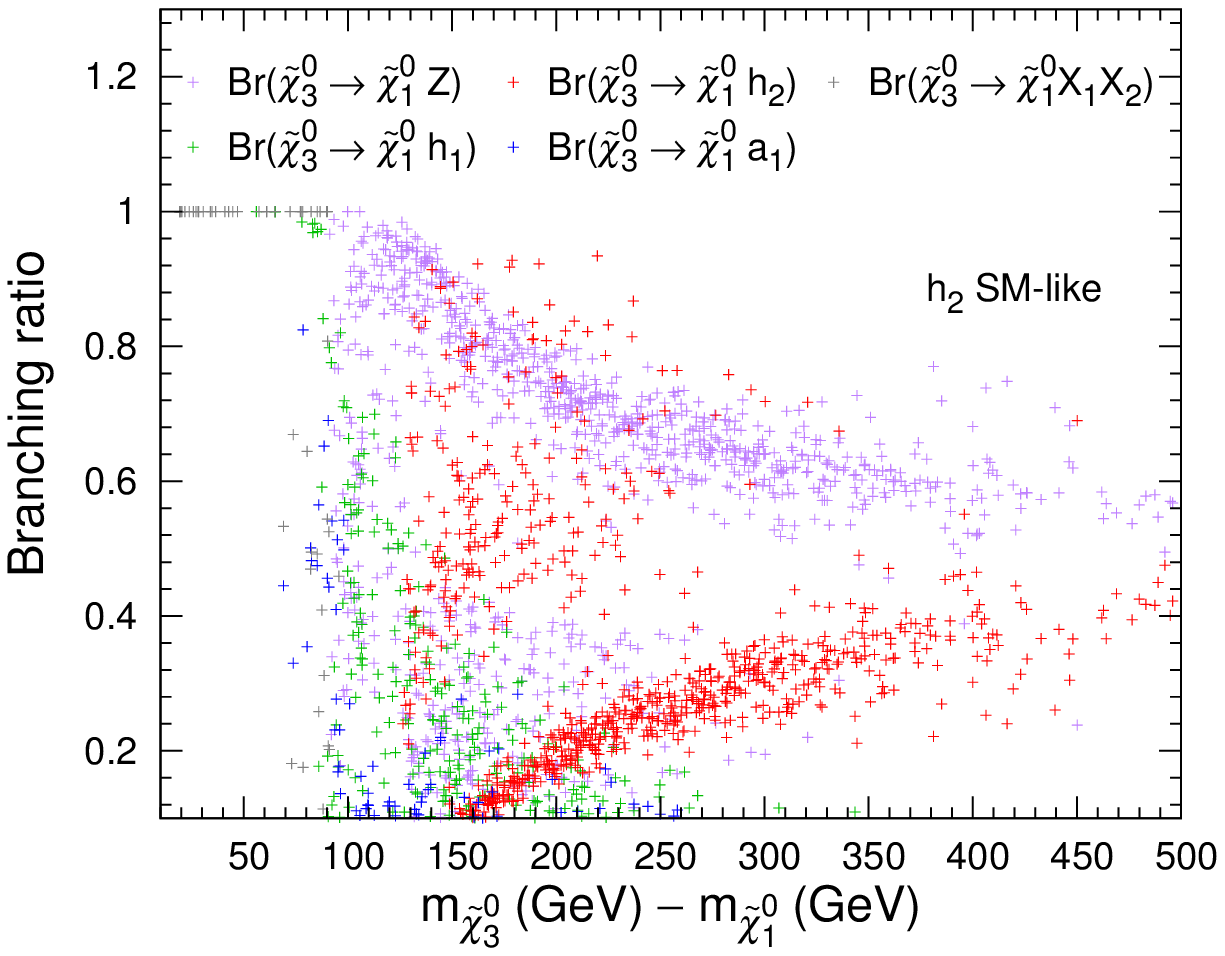}}
\caption{Decay branching ratios of $\chib$ (\ref{fig:branch:a1}, \ref{fig:branch:a2}) and $\chic$ (\ref{fig:branch:b1}, \ref{fig:branch:b2}) for $h_1$ or $h_2$ is the SM-like Higgs boson.
Purple, green, red, and blue points correspond to decays into $\chia Z$, $\chia \ha$, $\chia \hb$, and $\chia \aa$, respectively.
Gray points represent 3-body decay branching ratios.
}
\label{fig:branch}
\end{figure}

In Fig.~\ref{fig:branch} we present branching ratios of $\chib$ and $\chic$ decaying into $\chia$ versus mass differences $m_{\chib}-m_{\chia}$ and $m_{\chic}-m_{\chia}$, respectively.
These ratios affect the LHC discovery possibility of the parent particles.
Here we illustrate four typical decay channels, $\chia Z$, $\chia \ha$, $\chia \hb$, and $\chia \aa$.
We also show 3-body decay branching ratios of $\chib$ and $\chic$ in Fig. 2. These decay modes are typically dominant when $m_{\chibc} - m_{\chia} < m_Z$ for $\kappa/\lambda \gtrsim 0.4$
If the decay channels into $\chia$ and the SM-like Higgs are kinematically allowed, they would be sizable, and even dominant for $\chic$ decays.
$\br(\chibc \to \chia a_1)$ deceases as the mass differences increase, becoming negligible when $m_{\tilde\chi_{2,3}^0}-m_{\chia} \gtrsim 150~\GeV$.

We pick up three benchmark points to represent typical cases, as listed in Table~\ref{tab:bp}.
The dominant $\chib$ decay channel in BP1 is $\chib \to \chia Z$.
This is the most probable case, as we can see from Fig.~\ref{fig:branch:a2}.
On the other hand, $\chib$ in BP2 and BP3 mainly decays into $\chia h_1$ and $\chia a_1$, respectively, because the $\chia Z$ channel is kinematically forbidden.
In these benchmark points, $h_1$ and $a_1$ are almost pure singlet scalars.
$\chia$ in BP1 is almost pure singlino, but it could still effectively annihilate to give a correct relic abundance with its tiny higgsino components, due to the Higgs resonance enhancement.
Although the dominant decay channels of $\chib$ and $\chic$ in BP2 and BP3 are different ($\chia h_1$ and $\chia a_1$), their production signatures may be similar, as $h_1$ and $a_1$ have analogous decays ($\sim 92\%$ into $b\bar{b}$ and $\sim 7\%$ into $\tau^+\tau^-$).
In addition, $\chiapm$ in BP2 and BP3 can only decay into off-shell $W$ bosons, leading to softer visible products compared with BP1.

\begin{table}[htb]
\centering
\setstretch{1.2} 
\setlength\tabcolsep{0.4em}
\begin{tabular}{c|c|c|c}
\hline\hline
      & BP1 & BP2 & BP3   \\
\hline
  $\lambda$, $\kappa$      & 0.091, 0.016  & 0.270, 0.100  & 0.368, 0.144 \\
  $\tanb$, $\mueff$ (GeV)  & 39.6, 163.3   & 35.1, 121.3  & 35.6, 121.0 \\
  $A_\kappa$ (GeV), $A_\lambda$ (TeV) & $-35.9$, $8.94$ & $-173.4$, $3.79$ & $-8.77$, $4.43$ \\
\hline
$m_{\chia}$  (GeV) & 59.6 & 77.0 & 71.7 \\
$m_{\chib},~m_{\chic},~m_{\chiapm}$  (GeV) & 169, 173, 170  & 134, 146, 126 & 137, 160, 126 \\
$m_{\ha},~m_{\hb},~m_{\aa}$  (GeV) &46.0, 126, 55.8 & 23.0, 125, 153 &95.3, 125, 38.7 \\
$|N_{13}|^2+|N_{14}|^2$, $|N_{15}|^2$   &1.3\%, 98.7\% & 33.2\%, 66.8\% &43.5\%, 56.4\% \\
\hline
$\Omega_{\chia}h^2$ & 0.120 & 0.059 & 0.067 \\
\hline
\multirow{2}{*}{$\br(\chib \to \chia X)$ } & \multirow{2}{*}{$Z$ 98.7\%}  & $\ha$ 84.4\%, $q\bar{q}$ 10.6\%
& \multirow{2}{*}{$\aa$ 98.6\%} \\
&  & $l\bar{l}$ 3\%, $v_l \bar{v}_l$ 3\% & \\
 \hline
\multirow{2}{*}{$\br(\chic \to \chia X)$ } & $Z$ 97.1\%  & \multirow{2}{*}{$\ha$ 100\%} & $\aa$ 73.2\%, $q\bar{q}$ 14\% \\
& $\aa$ 2.7\%  & & $l\bar{l}$ 2\%, $v_l \bar{v}_l$ 4\% \\
\hline
\multirow{2}{*}{$\br(\ha/a_1 \to b\bar{b}/\tau^+ \tau^-)$} & \multirow{2}{*}{/} &$\ha \to b\bar{b}$ 91.8\%  & $\aa \to b\bar{b}$ 91.8\% \\
                 &                     & $\ha \to \tau^+ \tau^-$ 7.3\% & $\aa \to \tau^+ \tau^-$ 7.7\% \\
\hline\hline
\end{tabular}
\caption{Information of benchmark points.
$\chia X$ means $\chia$ associated with the particle(s) indicated in the entries.
$q\bar{q}$, $l\bar{l}$, and $v_l\bar{v}_l$ represent the sums over quarks, charged leptons, and neutrinos, respectively.}
\label{tab:bp}
\end{table}

\section{LHC searches}
\label{sec:LHC}

Compared with colored superpartners, the production of electroweak superpartners at the LHC yields much lower rates.
A helpful search strategy is to make use of $\ge 2$ charged leptons produced in decays of neutralinos, charginos, and sleptons.
SM backgrounds in these multilepton channels are quite clean.

With an integrated luminosity of $\sim 20~\ifb$ at the $8~\TeV$ LHC, both the ATLAS and CMS collaborations reported their search results for MSSM charginos and neutralinos in the $3l + \missET$~\cite{Aad:2014nua,Khachatryan:2014qwa} and $2l + \missET$~\cite{Aad:2014vma} final states.
The $3l + \missET$ search is particularly sensitive to $\chiapm \chibc$ production, which is a major process of electroweak SUSY production.
Assuming $\chiapm$ and $\chib$ are both pure wino with $\br(\chiapm \to \chia W^{\pm(*)}) = \br(\chib \to \chia Z^{(*)}) = 100 \%$, the ATLAS analysis has excluded $m_{\chiapm}$ and $m_{\chib}$ up to $\sim 350~\GeV$ at 95\% CL.
The $2l + \missET$ channel can be used to search for the $\tilde{\chi}_1^+\tilde{\chi}_1^-$ production, but it is less sensitive, just excluding $m_{\chiapm}$ up to $\sim 180~\GeV$ at 95\% CL.

In the singlino-higgsino scenario, $\chiapm$, $\chib$, and $\chic$ are higgsino-dominated.
We consider the production processes $ pp \to \chibc \chibc$, $\chibc \chiapm$, and $\tilde{\chi}_1^{+} \tilde{\chi}_1^{-}$ at the LHC.
$\chib$ and $\chic$ can decay into $\chia Z^{(\ast)}$, $\chia h_{1,2}^{(\ast)}$, and $\chia a_1^{(\ast)}$, while $\chiapm $ basically decays into $\chia W^{\pm (\ast)}$.
Because the doublet (higgsino) coupling to $W$ is weaker than the triplet (wino) coupling, the production rates of $\chiapm \chib$ and $\chiapm \chic$ here are much lower than the $\chiapm \chib$ production rate in the pure wino case.
Moreover, decays into $\chia$ and a Higgs scalar cannot be neglected and they hardly contribute to the trilepton final state as the scalar mainly decays into $b\bar{b}$.
Therefore, constraints from the $3 l + \missET$ searches are expected to be weaker than the pure wino case.
This situation is similar to that in the bino-higgsino scenario~\cite{Calibbi:2014lga}.

In order to evaluate the current constraints, we recast the ATLAS $3 l + \missET$~\cite{Aad:2014nua} and $2l + \missET$~\cite{Aad:2014vma} analyses to the singlino-higgsino scenario based on a Monte Carlo simulation.
In the simulation, we use \texttt{MadGraph~5}~\cite{Alwall:2014hca} to generate background and signal samples, and use \texttt{PYTHIA~6}~\cite{Sjostrand:2006za} to deal with the parton shower, particle decay, and hadronization processes.
The MLM scheme~\cite{Mangano:2006rw} is employed to handle the matching between matrix element and parton shower calculations.
\texttt{Delphes~3}~\cite{deFavereau:2013fsa} is utilized to carry out a fast detection simulation with the ATLAS setup.
Jets are clustered using the anti-$k_\mathrm{T}$ algorithm~\cite{Cacciari:2008gp} with a radius parameter of $R = 0.4$.

We generate simulation samples for $\chibc \chibc$, $\chibc \chiapm$, and $\tilde{\chi}_1^{+} \tilde{\chi}_1^{-}$ production and apply the same cuts in various signal regions of the ATLAS analyses.
The exclusion limits projected in the $m_{\chia}$-$m_{\chib}$ plane are shown in Fig.~\ref{fig:mchi}, where the red points are excluded at 95\% C.L.
Since both $\chib$ and $\chiapm$ are almost pure higgsinos, their masses are close, determined by $\mueff$.
We find that the ATLAS $3l + \missET$ searches have excluded $m_{\chib,\chiapm}$ up to $\sim 250~\GeV$, which is roughly 100~GeV lower than the pure wino case~\cite{Aad:2014nua}.
The $2l + \missET$ constraints would be even weaker.
Below we turn to evaluating the LHC sensitivities at $\sqrt{s}= 13~\TeV$ and 14~TeV.

\begin{figure}[!htbp]
\centering
\subfigure[~$3l + \missET$ channel\label{fig:mchi:3l}]{
\includegraphics[width=0.45\textwidth]{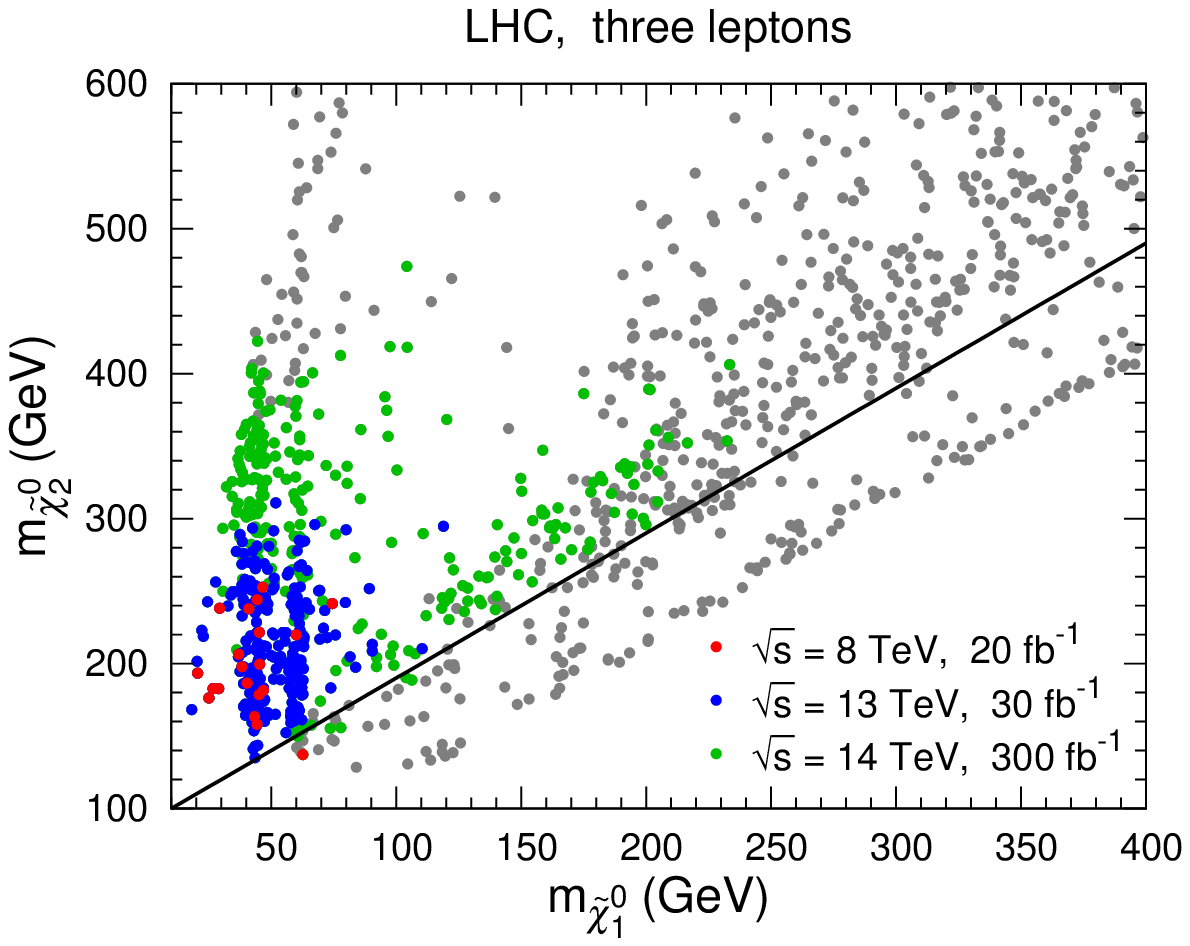}}
\subfigure[~$2l + \missET$ channel\label{fig:mchi:2l}]{
\includegraphics[width=0.45\textwidth]{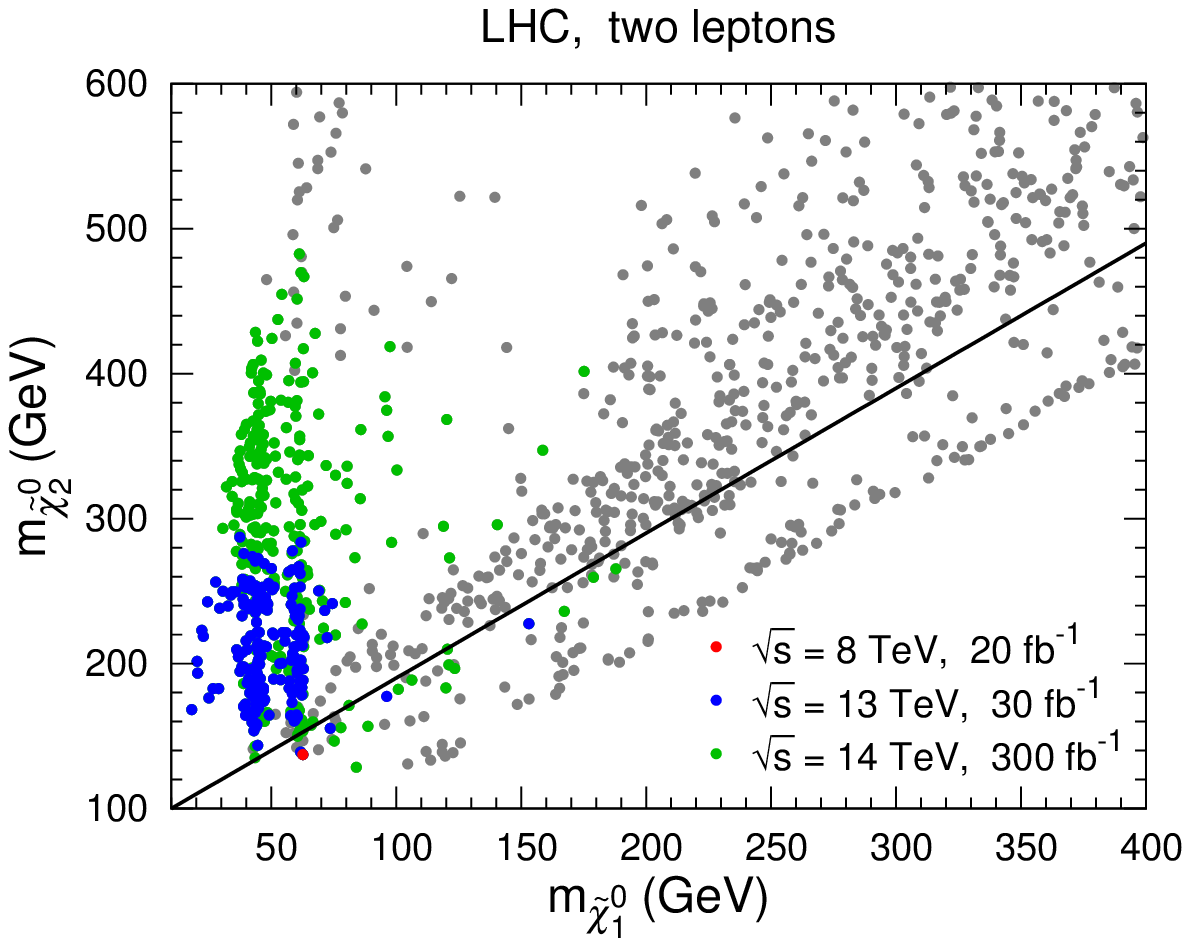}}
\caption{95\% CL exclusion results of the $3l + \missET$ (a) and $2l + \missET $ (b) searches in the $m_{\chia}$-$m_{\chib}$ plane.
Red points are excluded by the $8~\TeV$ ATLAS analyses with $\sim 20~\ifb$ data.
Blue (green) points are expected to be excluded by the $13~\TeV$ ($14~\TeV$) LHC with an integrated luminosity of $30~\ifb$ ($300~\ifb$).
Gray points will survive from the above searches.
The solid black lines denote the threshold $m_{\chib} = m_{\chia} + m_Z$.
}
\label{fig:mchi}
\end{figure}

\subsection{Prospect in the $3l + \missET$ channel}

In the $3l + \missET$ search channel, dominant SM backgrounds are $WZ$ and $ZZ$ production.
Minor backgrounds include $t\bar{t}$, $t \bar{t} V$ ($V=W,Z$), $t Z$, $VVV$, and Higgs production and so on.
We will omit these minor backgrounds for simplicity.
In order to efficiently suppress backgrounds and increase the signal significance, we adopt the following selection cuts. Hereafter a charged lepton $l$ denotes an electron or a muon.
\begin{description}
\item[Basic cuts]
select the events with exact three charged leptons which satisfy $\pT > 20~\GeV$ and $|\eta| < 2.5$ and are separate from each other by $\Delta R > 0.3$;
veto the events containing a $b$-jet with $\pT > 30~\GeV$ and $|\eta| < 2.5$;
select the events with $|\msfos - m_Z| < 10~\GeV$.
\item[$\missET$ cut]
select the events with $\missET > 50~\GeV$ or $100~\GeV$.
\item[$\mT$ cut]
select the events with $\mT > 100~\GeV$.
\end{description}
Here $\msfos$ is the invariant mass of a same-flavor opposite-sign (SFOS) lepton pair. When there are two such pairs, we choose the one with an invariant mass closer to $m_Z$.
Events without an SFOS pair are discarded.
$\mT$ is the transverse mass defined as $\mT = \sqrt{ 2 (p_\mathrm{T}^l \missET - \mathbf{p}_\mathrm{T}^{l} \cdot \mathbf{p}_\mathrm{T}^\mathrm{miss})}$, where $\mathbf{p}_\mathrm{T}^\mathrm{miss}$ is the missing transverse momentum vector and the lepton $l$ is the one not forming the SFOS lepton pair.
For the $\missET$ cut, we adopt two thresholds, 50~GeV and 100~GeV, optimized for light and heavy $m_{\chib,\chiapm}$, respectively.

\begin{figure}[!htbp]
\centering
\subfigure[~$\msfos$ distributions]{
\includegraphics[width=0.45\textwidth]{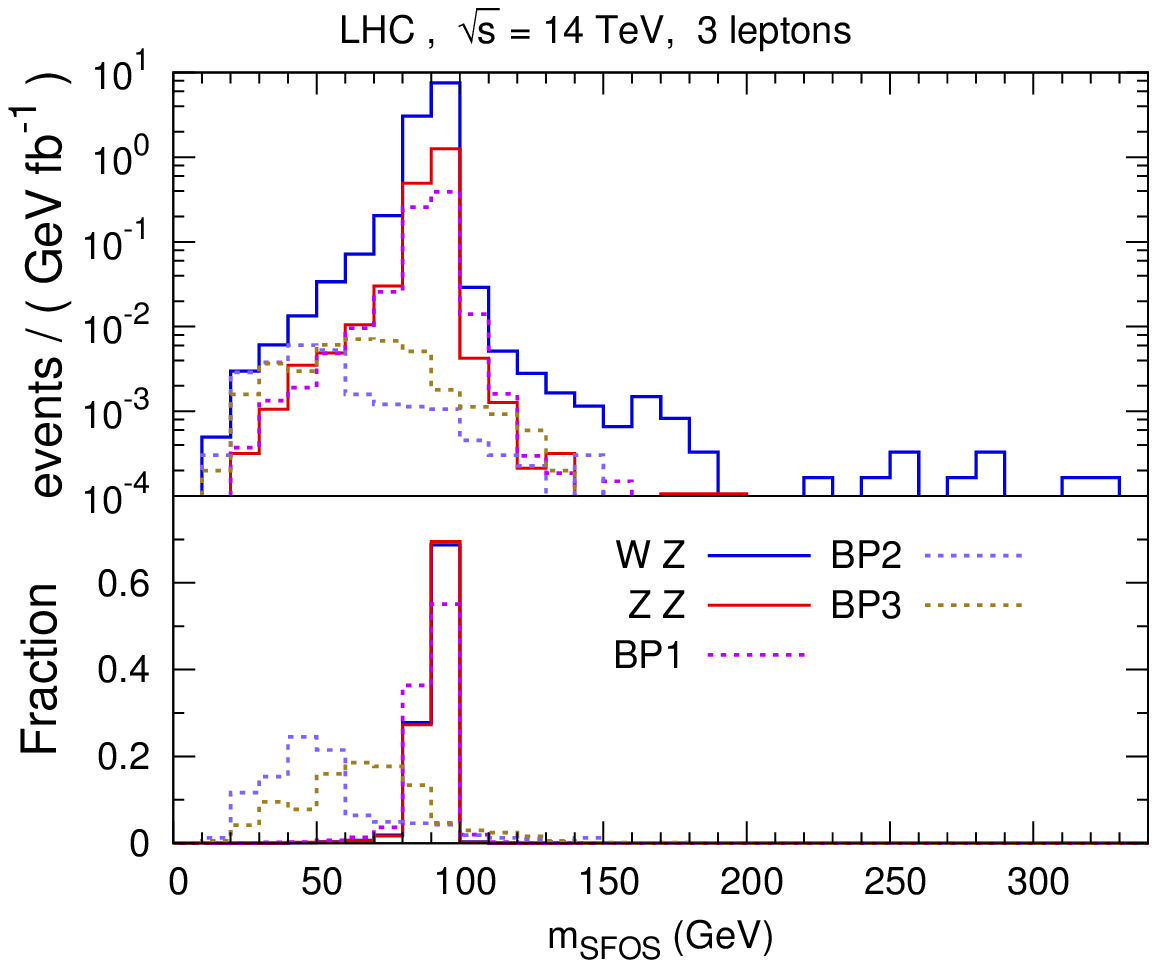}}
\subfigure[~$\missET$ distributions]{
\includegraphics[width=0.45\textwidth]{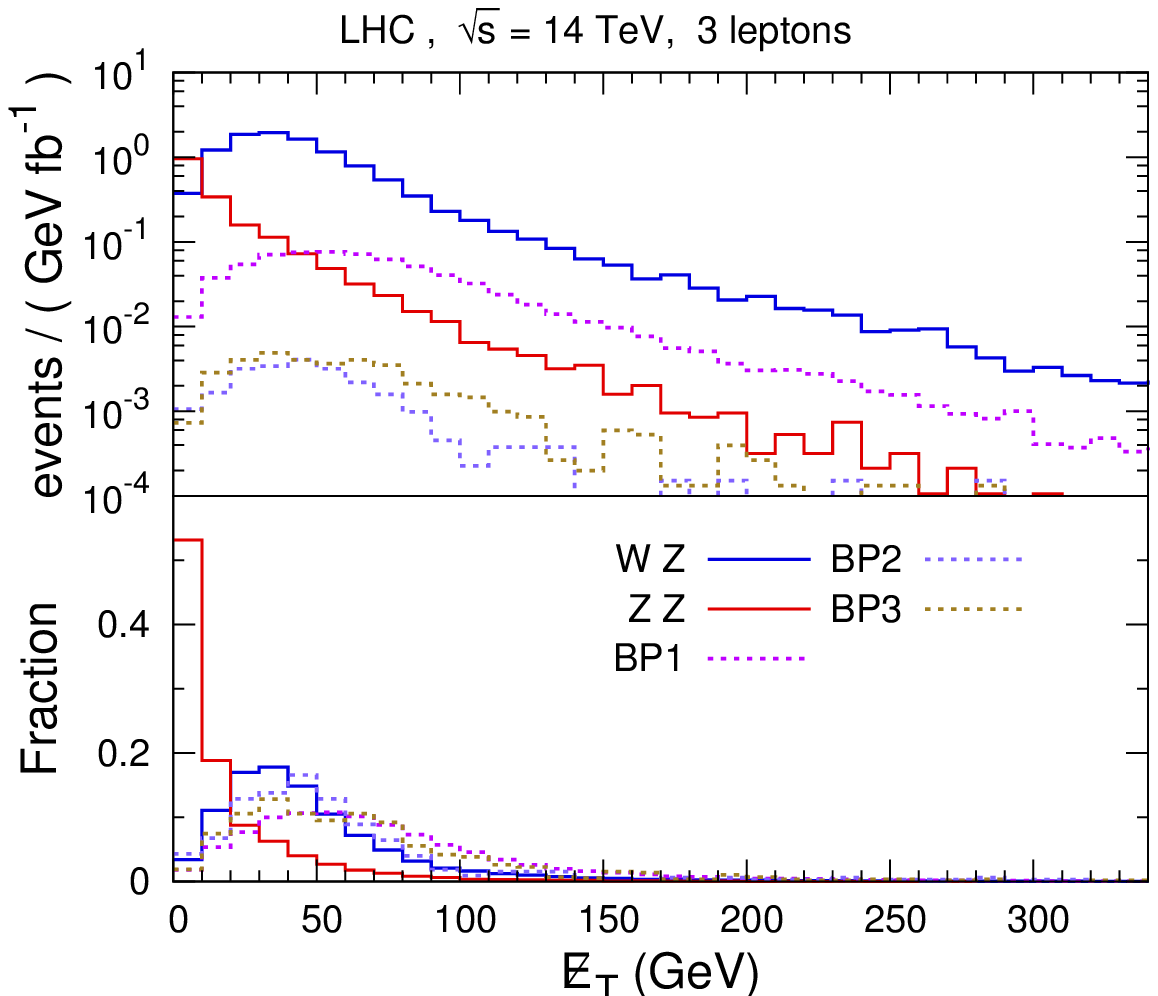}}
\subfigure[~$\mT$ distributions]{
\includegraphics[width=0.45\textwidth]{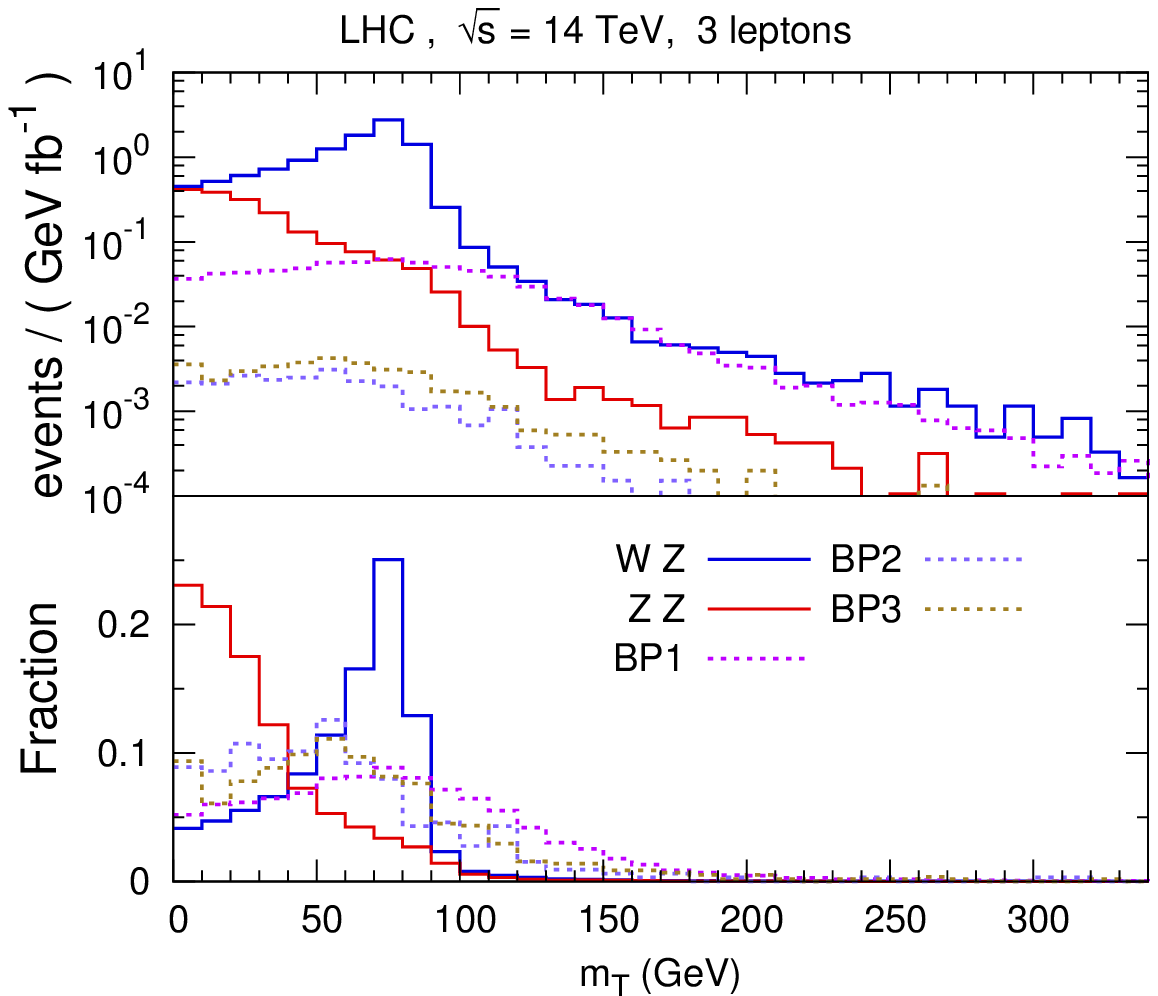}}
\caption{$\msfos$ (a), $\missET$ (b), and $\mT$ (c) distributions for backgrounds and signal benchmark points in the $3l + \missET$ channel at the 14~TeV LHC.}
\label{fig:3l_bmp}
\end{figure}

In Fig.~\ref{fig:3l_bmp}, we demonstrate the $\msfos$, $\missET$, and $\mT$ distributions of backgrounds and signals after the basic cuts except $|\msfos - m_Z| < 10~\GeV$.
The $m_\mathrm{SFOS}$ variable is chiefly used to reconstruct $Z$ bosons from their $l^+l^-$ products. Therefore, there is a clear peak near $m_Z$ in the $m_\mathrm{SFOS}$ distributions for $WZ$ and $ZZ$, as well as that for the signal BP1 where both $\chib$ and $\chic$ dominantly decay into $\chia Z$.
On the other hand, both $\chib$ and $\chic$ in BP2 and BP3 primarily decay into $\chia$ and a Higgs boson ($h_1$ or $a_1$), which subsequently decay into $\tau^+\tau^-$ with a branching ratio lower than 10\%.
Thus, the peaks in the $m_\mathrm{SFOS}$ distributions for BP2 and BP3 are not at $m_Z$.
One reason for this is that the relevant decay products $h_1$ and $a_1$ are typically lighter than $Z$.
Another one is that electrons and muons from tau leptonic decays has lower energies due to the associated neutrinos.
Therefore, the cut condition $|\msfos - m_Z| < 10~\GeV$ is only optimized for the case like BP1.

The $3l$ final state from the $ZZ$ background mainly comes from the case that both $Z$ bosons decay into $l^+l^-$ pairs but one lepton cannot be successfully reconstructed. In this case there is no neutrino contributing $\missET$.
Thus its $\missET$ distribution is softer than others, and so is its $m_\mathrm{T}$ distribution.
For the $WZ$ background, the $\mT$ variable is bounded by the $W$ boson mass, hence the distribution has an obvious endpoint near $m_W$.

\begin{table}[htb]
\centering
\setstretch{1.2} 
\setlength\tabcolsep{0.4em}
\begin{tabular}{ccccccccc}
\hline\hline
& $WZ$   & $ZZ$            & \multicolumn{2}{c}{BP1}  & \multicolumn{2}{c}{BP2} & \multicolumn{2}{c}{BP3}  \\
& $\sigma$ & $\sigma$ &$\sigma$ & $\mathcal{S}$   &$\sigma$ & $\mathcal{S}$ &$\sigma$ & $\mathcal{S}$   \\
\hline
Basic cuts & 105  &17.3   & 6.39  & 0.52  & 0.021 &0.0017	 &0.060 &0.0049\\
$\missET > 50~\GeV~$&37.2 &1.51  & 4.11  & 1.06 &0.008  &0.0021 &0.034 &0.0087\\
$\mT > 100~\GeV$& 1.22 & 0.06 &1.60 &9.93    &0.004  &0.0278 &0.014 &0.0973\\
\hline\hline
\end{tabular}
\caption{Visible cross sections $\sigma$ (in fb) for backgrounds and signal benchmark points after each cut in the $3l+\missET$ channel at the 14~TeV LHC. The signal significances ($\mathcal{S}$) assuming an integrated luminosity of $300~\ifb$ are also listed.}
\label{tab:bp_3l}
\end{table}

Table~\ref{tab:bp_3l} lists visible cross sections of backgrounds and signals as well as signal significances assuming $\sqrt{s}=14~\TeV$ and an integrated luminosity of $300~\ifb$ after each cut.
Here the visible cross section is defined as the production cross section multiplied by the acceptance and efficiency.
The signal significance $\mathcal{S}$ is defined as $S/\sqrt{S+B+(0.1 B)^2}$, with $S$ ($B$) denoting the event number of signals (backgrounds).
A $10\%$ systematic uncertainty on the backgrounds has been considered in our analysis.
We find that the $\missET > 50~\GeV$ and $\mT >100~\GeV$ cuts suppress the $WZ$ ($ZZ$)  background by 2 ($3$) orders of magnitude.
Consequently, the signal significance for BP1 is efficiently increased.
It is expected to reach the $9.9\sigma$ significance with a data set of $300~\ifb$.

The expected 95\% CL exclusion results at the $\sqrt{s}=13~\TeV$ and 14~TeV have been presented in Fig.~\ref{fig:mchi:3l}.
With an integrated luminosity of $30~(300)~\ifb$, LHC searches are expected to reach up to $m_{\chib,\chiapm} \sim 320~(420)~\GeV$.
There are many points with $m_{\chib} \lesssim m_{\chia} + m_Z$ may not be explored even with a data set of $300~\ifb$ at the 14~TeV LHC, because $\chib$ may have a small $\chia Z$ branching ratio or decay into an off-shell $Z$ boson.

\subsection{Prospect in the $2l+\missET$ channel}

Major backgrounds in the $2l+\missET$ channel are $WW$, $WZ$, $ZZ$, and $t\bar{t}$ production. The following selection cuts are used.
\begin{description}
\item[Basic cuts]
select the events with exact two opposite-sign charged leptons which satisfy $\pT > 20~\GeV$ and $|\eta| < 2.5$;
the harder lepton should have $\pT > 30~\GeV$;
if the two leptons are the same flavor, their invariant mass should satisfy $\msfos > 20~\GeV$ and $|\msfos - m_Z| > 10~\GeV$.
\item[Jet veto]
veto the events containing any jet with $\pT > 30~\GeV$ and $|\eta| < 2.5$.
\item[$\mTT$ cut]
select the events with $\mTT > 90~\GeV$, $120~\GeV$, or $150~\GeV$.
\end{description}
Note that the condition $\msfos > 20~\GeV$ is used to avoid low mass hadronic resonances.
$\mTT$ is defined as \cite{Lester:1999tx,Barr:2003rg}
\begin{equation}
  \mTT = \underset{\mathbf{p}_\mathrm{T}^1 + \mathbf{p}_\mathrm{T}^2 =  \mathbf{p}_\mathrm{T}^\mathrm{miss}}{\mathrm{min}}
  \{\mathrm{max}[m_\mathrm{T}(\mathbf{p}_\mathrm{T}^a,\mathbf{p}_\mathrm{T}^1), m_\mathrm{T}(\mathbf{p}_\mathrm{T}^b,\mathbf{p}_\mathrm{T}^2) ] \} ,
  \label{eq:mT2}
\end{equation}
where $m_\mathrm{T}(\mathbf{p}_\mathrm{T}^i,\mathbf{p}_\mathrm{T}^j)=\sqrt{2(p_\mathrm{T}^i p_\mathrm{T}^j
-\mathbf{p}_\mathrm{T}^i\cdot\mathbf{p}_\mathrm{T}^j)}$, and $\mathbf{p}_\mathrm{T}^a$ and $ \mathbf{p}_\mathrm{T}^b$ are the transverse momenta of two visible particles in the decay chain, which are the two leptons in our case.
$\mathbf{p}_\mathrm{T}^1$ and $\mathbf{p}_\mathrm{T}^2$ are a partition of the missing transverse momentum $\mathbf{p}_\mathrm{T}^\mathrm{miss}$.
As $\mTT$ is the minimum of the larger $m_\mathrm{T}$ over all partitions, its distribution for a pair production process with two semi-invisible decay chains has an upper endpoint, which is determined by the mass difference between the parent particle and its invisible child.
We use three thresholds for the $\mTT$ cut, aiming at varied mass splittings between $\chiapm$ and $\chia$.

\begin{figure}[!htbp]
\centering
\subfigure[~$\msfos$ distributions~\label{fig:2l_bmp:mll}]{
\includegraphics[width=0.45\textwidth]{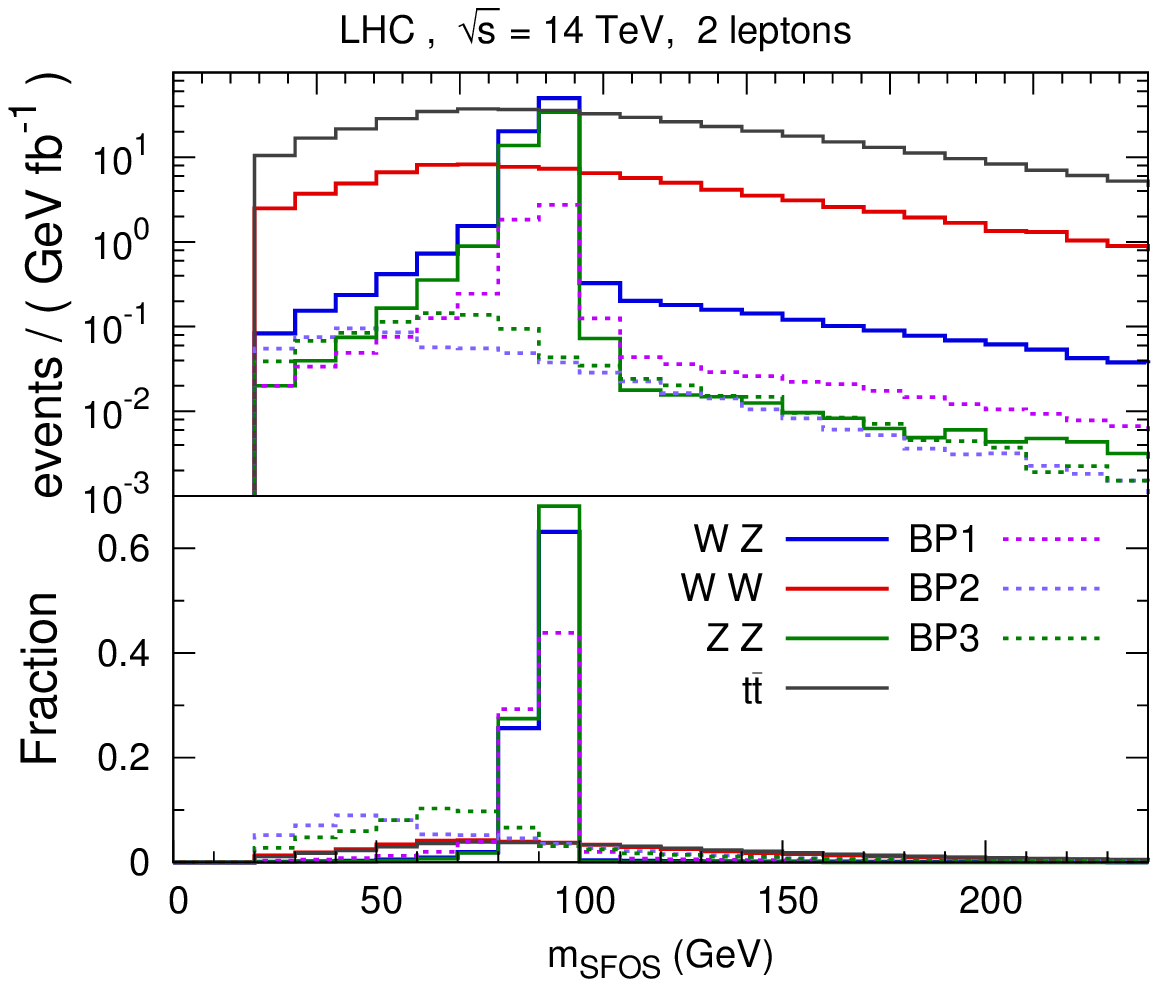}}
\subfigure[~$\mTT$ distributions~\label{fig:2l_bmp:mT2}]{
\includegraphics[width=0.45\textwidth]{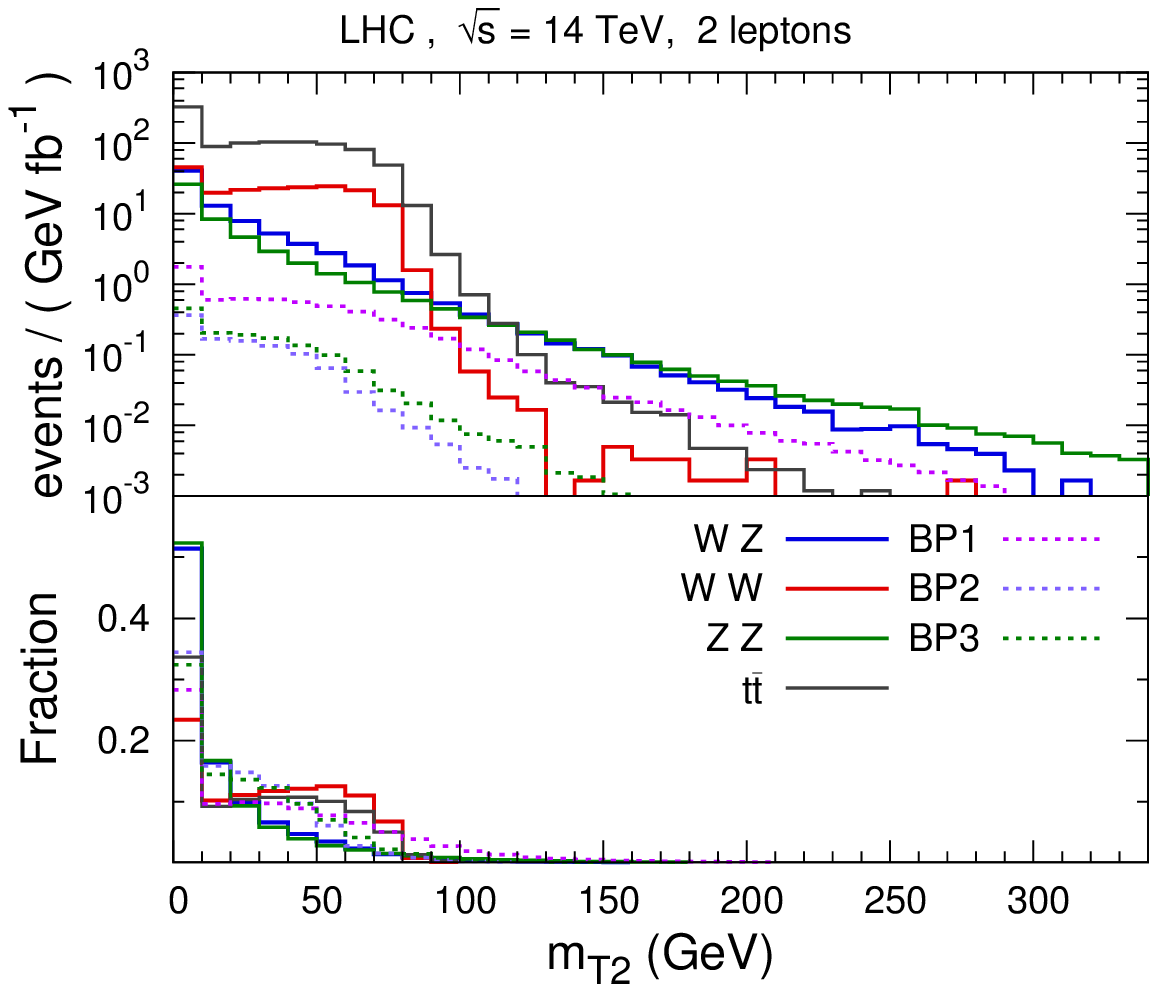}}
\caption{$\msfos$ (a) and $\mTT$ (b) distributions for backgrounds and signal benchmark points in the $2l + \missET$ channel at the 14~TeV LHC.
}
\label{fig:2l_bmp}
\end{figure}

Fig.~\ref{fig:3l_bmp} shows the $\msfos$ and $\mTT$ distributions after the basic cuts except $|\msfos - m_Z| > 10~\GeV$.
For the $ZZ$ and $WZ$ backgrounds, there can be a SFOS lepton pair induced by one $Z$ boson.
This leads to peaks around $m_Z$ in the $\msfos$ distributions, which are distinct in Fig.~\ref{fig:2l_bmp:mll}.
The condition $|\msfos - m_Z| > 10~\GeV$ aims at excluding such events.
For BP1, these is also a peak around $m_Z$ induced by $\chiapm\chib$ production, which, however, is not the target of the $2l+\missET$ search.
As illustrated in Fig.~\ref{fig:2l_bmp:mT2}, the $\mTT$ distributions for the $WW$ and $t\bar{t}$ backgrounds are essentially bounded by $m_W$,
while that for BP1 extends to higher values.

\begin{table}[htb]
\centering
\setstretch{1.2} 
\setlength\tabcolsep{0.4em}
\begin{tabular}{ccccccccccc}
\hline\hline
& $WZ$        & $ZZ$     & $WW$ & $t\bar{t}$        & \multicolumn{2}{c}{BP1}  & \multicolumn{2}{c}{BP2} & \multicolumn{2}{c}{BP3}  \\
& $\sigma$ & $\sigma$ &$\sigma$  &$\sigma$    & $\sigma$ & $\mathcal{S}$ &$\sigma$ & $\mathcal{S}$ &$\sigma$ & $\mathcal{S}$
\\
\hline
Basic cuts          &88.8   &22.3  &1798  &8930     &16.8  &0.015  &9.75  &0.009  & 12.7   & 0.012\\
Jet veto           &35.8   &7.25   &848   &253      &8.23  &0.072  &5.42  &0.047  & 6.86   & 0.060\\
$\mTT > 90~\GeV$   &0.24   &0.32   &0.48  &0.98     &0.58  &2.608  &0.05  &0.229  & 0.13   & 0.594\\
\hline\hline
\end{tabular}
\caption{Visible cross sections $\sigma$ (in fb) and signal significances ($S$) after each cut in the $2l + \missET$ channel at the 14 TeV LHC.
The signal significances correspond to an integrated luminosity of $300~\ifb$.}
\label{tab:bp_2l}
\end{table}

Table~\ref{tab:bp_2l} demonstrates visible cross sections and signal significances after each cut at the 14~TeV LHC.
Because $b$-jets are always produced associating with the two leptons in the $t\bar{t}$ background, the veto on jets kills $\sim 97\%$ events of this background.
The $\mTT > 90~\GeV$ cut is pretty powerful in suppressing the $WW$ and $t\bar{t}$ backgrounds, reducing them by $2-3$ orders of magnitude.
Through these cuts, the significance of BP1 reaches above $2.6\sigma$ for $300~\ifb$ of data.
For BP2 and BP3, $m_{\chiapm} - m_{\chia} < m_W$, leading to $m_{\chiapm}$ decays into off-shell $W$ bosons and hence soft $\mTT$ distributions.
Although the $\mTT$ cut seems to discard some signal events, this condition is necessary.
If not, the Drell-Yan background $pp\to l^+l^-$ would be enormous, because we have not included a $\missET$ cut in the basic cuts. Actually, the $\mTT$ cut here also serves as a $\missET$ cut.

The expected exclusion on the parameter points at $\sqrt{s}=13~\TeV$ and 14~TeV has been shown in Fig.~\ref{fig:mchi:2l}.
With an integrated luminosity of $30~(300)~\ifb$, the LHC $2l+\missET$ search could reach up to $m_{\chib,\chiapm} \sim 280~(480)~\GeV$.
However, many parameter points with $m_{\chiapm} - m_{\chia} \lesssim m_W$, as well as BP2 and BP3, will not be able to be probed because their $\mTT$ distributions cannot extend much beyond $m_W$.

\section{Direct and indirect detection}
\label{sec:exps}

In this section, we investigate the constraints from DM direct and indirect searches on the singlino-higgsino scenario, as well as the sensitivity of future experiments.

\subsection{Direct detection}

DM direct detection experiments search for recoil signals of target nuclei scattered off by incident DM particles. DM-nucleus scatterings can be classified into two types, spin-independent (SI) and spin-dependent (SD).
The SI scattering cross section is coherently enhanced by the square of the nucleon number in the nucleus.
SD scatterings have no such enhancement and depend on the particular spin property of the target nucleus.
Therefore, current direct detection experiments are much more sensitive to SI scatterings than SD scatterings.

\begin{figure}[!htbp]
\centering
\subfigure[~$m_{\chia}$-$\sigma_{p}^\mathrm{SI}$ plane\label{fig:SID:a}]{
\includegraphics[width=0.45\textwidth]{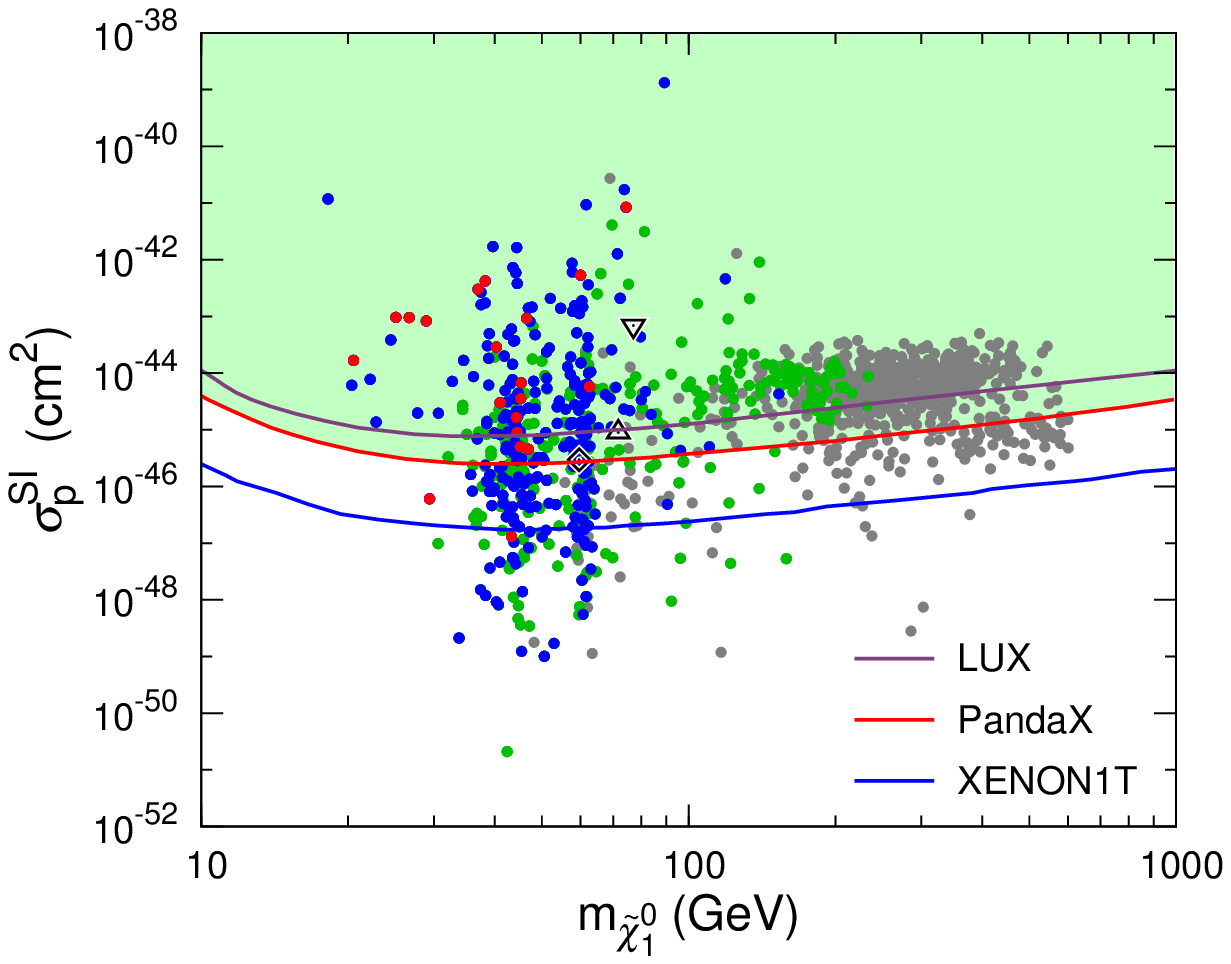}}
\subfigure[~$m_{\chia}$-$\xi \sigma_{p}^\mathrm{SI}$ plane\label{fig:SID:b}]{
\includegraphics[width=0.45\textwidth]{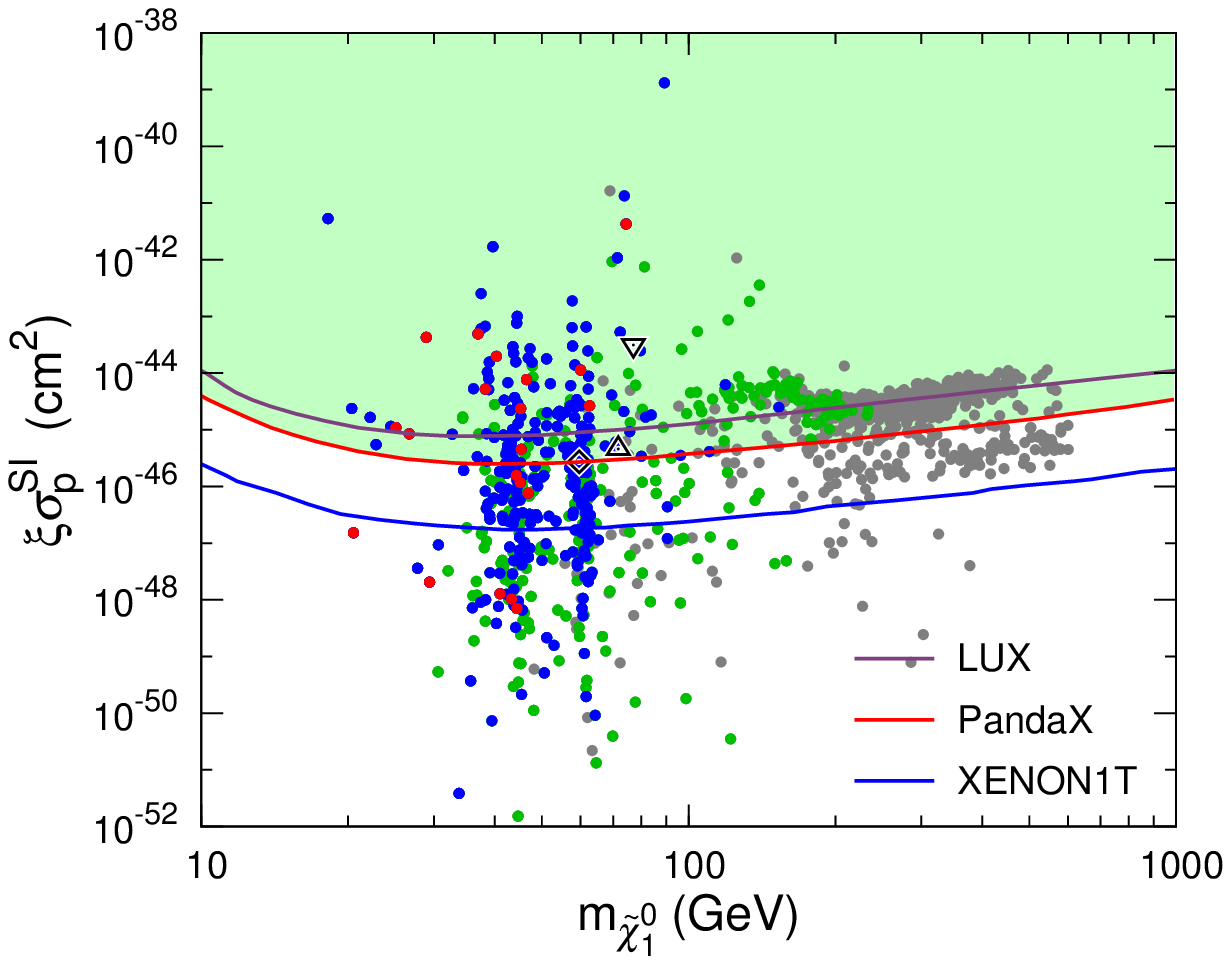}}
\subfigure[~$m_{\chia}$-$\sigma_{p}^\mathrm{SD}$ plane\label{fig:SID:c}]{
\includegraphics[width=0.45\textwidth]{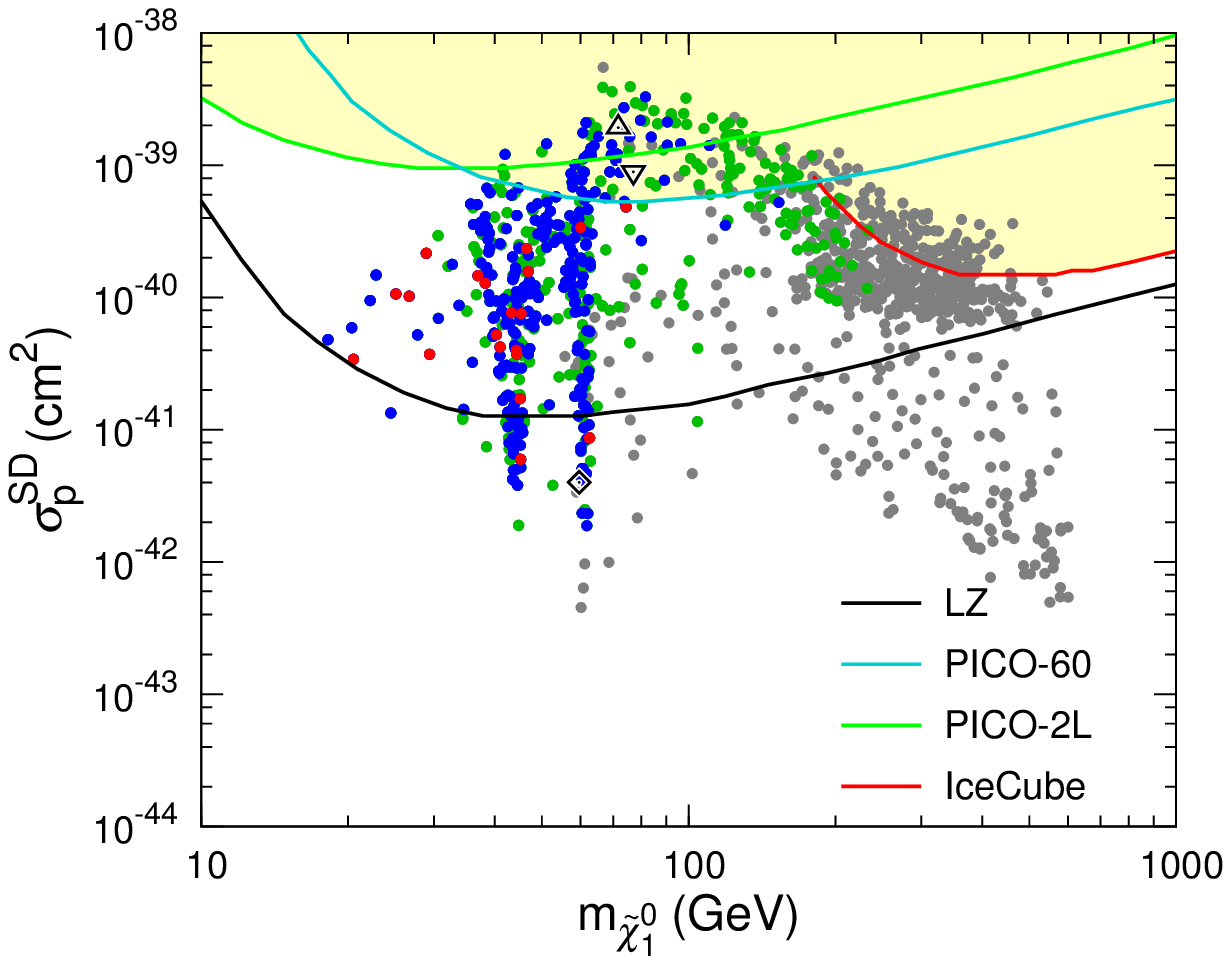}}
\subfigure[~$m_{\chia}$-$\xi \sigma_{p}^\mathrm{SD}$ plane\label{fig:SID:d}]{
\includegraphics[width=0.45\textwidth]{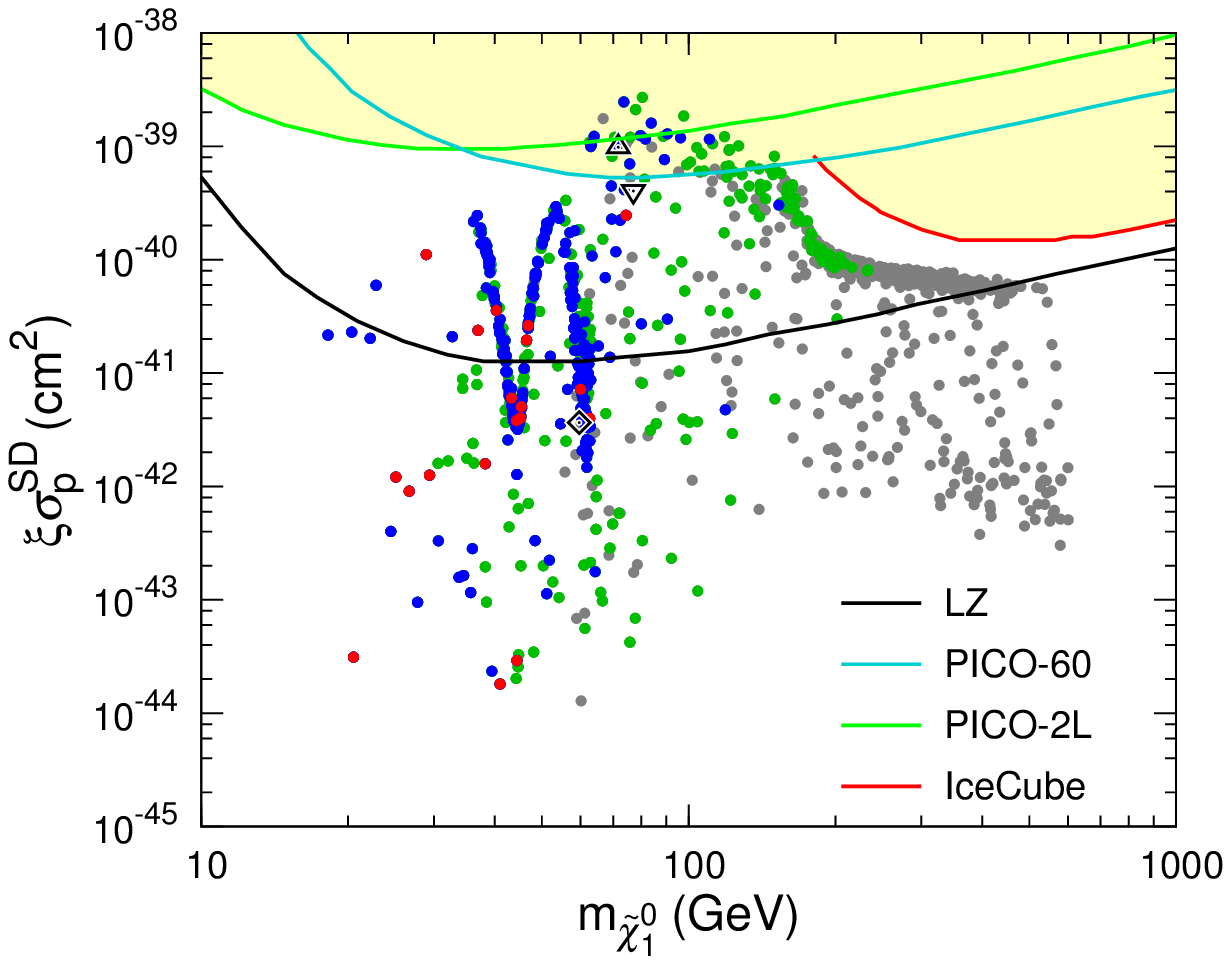}}
\caption{Parameter points projected into the $m_{\chia}$-$\sigma_{p}^\mathrm{SI}$ (a), $m_{\chia}$-$\xi \sigma_{p}^\mathrm{SI}$ (b),  $m_{\chia}$-$\sigma_{p}^\mathrm{SD}$ (c), and $m_{\chia}$-$\xi \sigma_{p}^\mathrm{SD}$ (d) planes.
The notation for the colored points is the same as in Fig.~\ref{fig:mchi},
while open diamonds, downward triangles, and upward triangles denote BP1, BP2, and BP3, respectively.
For the SI scattering, the exclusion limit from LUX~\cite{Akerib:2013tjd}, PandaX~\cite{Tan:2016zwf}, and the expected exclusion limit of XENON1T~\cite{Aprile:2015uzo} at 90\% CL are shown. For the SD scattering, the exclusion limits from PICO~\cite{Amole:2015lsj,Amole:2015pla} and IceCube~\cite{Aartsen:2016exj} and the expected exclusion limit of LZ~\cite{Akerib:2015cja} at 90\% CL are shown.
}
\label{fig:SID}
\end{figure}

Due to the Majorana nature, neutralino DM cannot have SI scatterings through the exchange of a $Z$ boson.
Nevertheless, in the singlino-higgsino scenario SI scatterings can be induced by the Higgs boson exchange, while SD scatterings by the $Z$ boson exchange.
The SI DM-proton cross section $\sigma_{p}^\mathrm{SI}$ and the SD DM-proton cross section $\sigma_{p}^\mathrm{SD}$ for the parameter points are shown in Figs.~\ref{fig:SID:a} and \ref{fig:SID:c}, respectively.
$\sigma_{p}^\mathrm{SD}$ is typically larger than $\sigma_{p}^\mathrm{SI}$ by $\sim 2-6$ orders of magnitude.

When $\Omega_{\chia} h^2 <0.107$, the possibility that just a faction of DM particles are contributed by $\chia$ should be taken into account.
For this reason, we introduce a density fraction defined by $\xi =\mathrm{min}(1,\Omega_{\chia} h^2/0.107)$.
In this case, the proper quantities for comparing with experimental results are the reduced SI and SD cross sections, $\xi\sigma_{p}^\mathrm{SI}$ and $\xi\sigma_{p}^\mathrm{SD}$, which are shown in Figs.~\ref{fig:SID:b} and \ref{fig:SID:d}, respectively.
As we can see from the left panel of Fig.~\ref{fig:omega}, the predicted relic density can be very low when $\chia$ could annihilate through the $Z$ or SM-like Higgs resonance.
Thus, $\xi$ can be as small as $\sim \mathcal{O}(10^{-3})$ and significantly reduce $\xi\sigma_{p}^\mathrm{SI}$ and $\xi\sigma_{p}^\mathrm{SD}$.

In Fig.~\ref{fig:SID:d}, some points  align as two curves reflecting the profiles of the $Z$ and SM-like Higgs resonances.
Since the singlino does not couple to $Z$, $\sigma_{p}^\mathrm{SD}$ is proportional to the higgsino components of $\chia$.
When the resonance enhancement works, $\Omega_{\chia} h^2$ is basically inversely proportional to the higgsino components in the $Z$ resonance case, as well as in the SM-like Higgs resonance case if the SM-like Higgs is doublet-dominated.
Consequently, $\xi\sigma_{p}^\mathrm{SD}$ can be a quantity independent of how large the higgsino components are, and hence reflects the resonance structure.
On the other hand, this behavior is not obvious in Fig.~\ref{fig:SID:b}, as $\sigma_{p}^\mathrm{SI}$ is generally determined by both the singlino and higgsino components.

In Figs.~\ref{fig:SID:a} and \ref{fig:SID:b}, we also plot the exclusion limit from LUX~\cite{Akerib:2013tjd}, PandaX~\cite{Tan:2016zwf}, and the projected exclusion limit for XENON1T~\cite{Aprile:2015uzo} in $2~\mathrm{t}\cdot\mathrm{year}$ exposure at $90\%$ CL for the SI scattering.
When the $\xi$ factor is not considered, the PandaX limit excludes a lot of parameter points, especially the bunch with $m_{\chia}\gtrsim 100~\GeV$.
After considering the $\xi$ factor, roughly a half of the points in this bunch can escape from the PandaX limit, as $\xi$ for them is typically $\sim \mathcal{O}(10^{-1})$.
Nevertheless, they will be covered in the XENON1T search.
BP2 and BP3 have already been excluded by the PandaX search, and it is quite promising to probe BP1 in the near future experiments.
When $\chia$ could annihilate through a resonance, no matter it is a $Z$, SM-like Higgs, or other Higgs resonance, an acceptable relic density and a small DM-nucleon scattering cross section could be simultaneously obtained.
In this case, there are many points that can evade the PandaX and XENON1T limits, but most of them would be well investigated in future LHC searches.
For the SD scattering, most stringent bounds come from the bubble chamber experiment PICO~\cite{Amole:2015lsj,Amole:2015pla}, as plotted in Figs.~\ref{fig:SID:c} and \ref{fig:SID:d}.
Although these bounds seem quite weak, they have excluded BP3.
The $90\%$ CL expected exclusion limit of LZ~\cite{Akerib:2015cja} in $5.6~\mathrm{t}\cdot 1000~\mathrm{day}$ exposure will cover down to the SD cross section of $\sim 10^{-41}~\cm^2$ and well investigate the singlino-higgsino scenario.

\subsection{Indirect detection}

As an independent approach to reveal the nature of DM, indirect detection experiments seek for high energy comic rays, gamma rays, and neutrinos induced by DM decays or annihilations in Galactic and extragalactic objects.
For $R$-parity conserved SUSY models, the LSP is absolutely stable.
Thus, indirect detection signatures come from LSP annihilation, which depends on the thermally averaged annihilation cross section $\sigmav$ and the DM density in annihilation regions.

As discussed in Sec.~\ref{sec:scan}, for $m_{\chia} \lesssim 70~\GeV$, a large $\sigmav$ at the freeze-out epoch is mainly achieved by the resonance enhancement of a $Z$ or Higgs boson.
However, the annihilation behavior at low velocities can be quite different and the cross section can be significantly suppressed. One reason for this is that the $s$-wave annihilation cross section into a fermion pair $f\bar{f}$ through an $s$-channel $Z$ is helicity suppressed and proportional to $m_f^2/m_{\chia}^2$.
Additionally, the leading order of annihilation through an $s$-channel CP-even Higgs is of $p$-wave.
Moreover, when annihilation into $f\bar{f}$ comes through an $s$-channel (CP-even or CP-odd) Higgs, the coefficient of any wave is proportional to $m_f^2/m_{\chia}^2$ due to the fermion couplings to neutral Higgs bosons.

\begin{figure}[!htbp]
\centering
\includegraphics[width=0.45\textwidth]{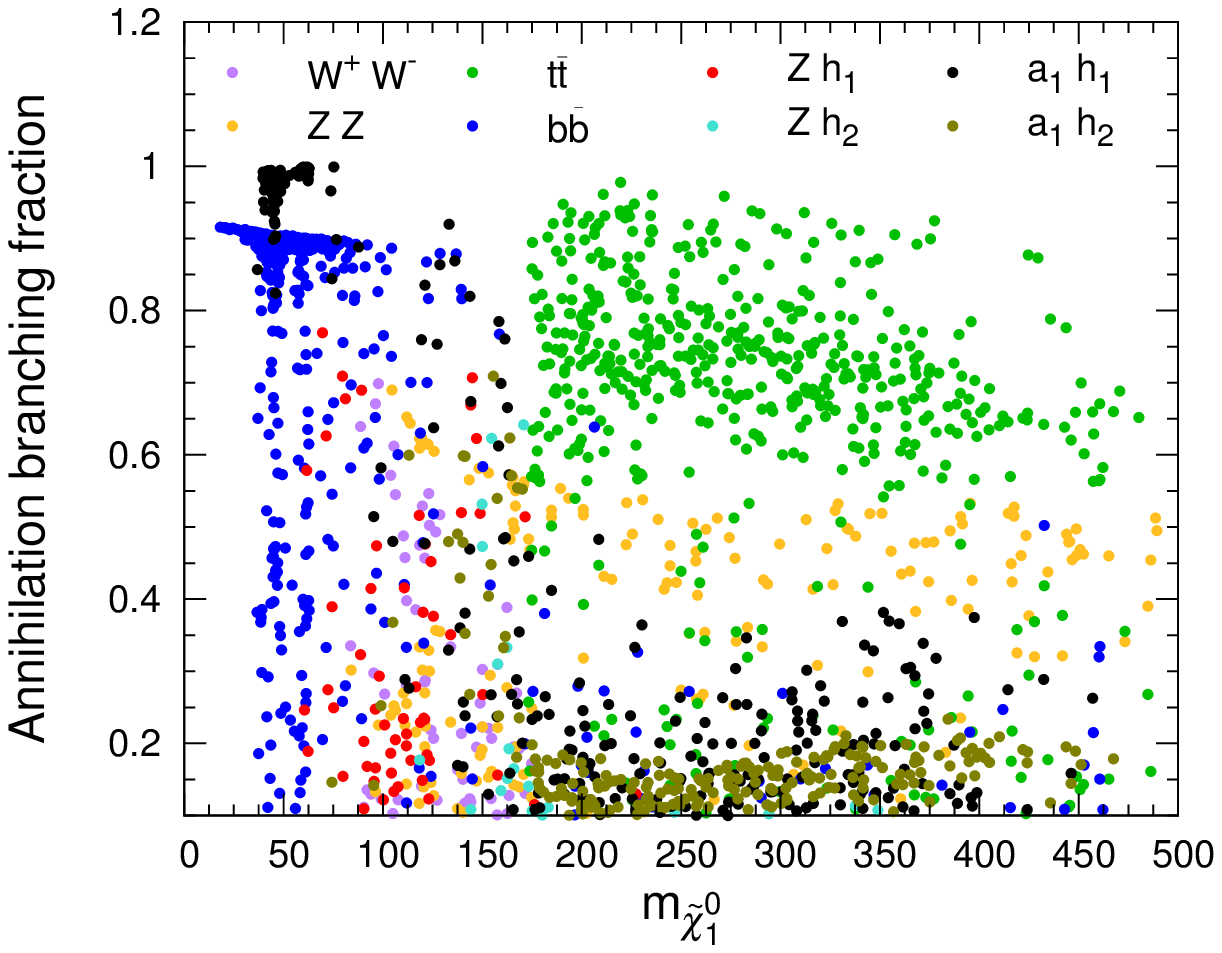}
\caption{Branching fractions of major $\chia\chia$ annihilation channels with $\sqrt{\left<v^2\right>} = 0.001$.}
\label{fig:anni_branch}
\end{figure}

Branching fractions of major annihilation channels for nonrelativistic DM with $\sqrt{\left<v^2\right>} = 0.001$ are shown in Fig.~\ref{fig:anni_branch}.
For $m_{\chia} < m_t$, the dominant annihilation channel is basically either $b\bar{b}$ or $a_1 h_1$.
When both $a_1$ and $h_1$ are light, the $a_1 h_1$ channel can be important at low velocities, although it could not compete with the $f\bar{f}$ channels at the freeze-out epoch.
This channel does not suffer from helicity suppression in contrast to $f\bar{f}$.
When the $a_1 h_1$ channel is not available, the $b\bar{b}$ channel would be the most important one because the $b$ quark is the heaviest SM fermion except the $t$ quark.
At the freeze-out epoch, if the $Z$ resonance enhancement was important, the annihilations into 5 types of light quarks were comparable to each other because $p$-wave annihilation was significant.
However, as the DM velocity goes down, the $b\bar{b}$ channel becomes dominant over others.

For $m_{\chia} > m_t$, the $t\bar{t}$ channel opens and becomes dominant.
It is less suppressed in $s$ wave, because $m_t$ has the same order of magnitude with the $m_{\chia}$ value we concern in this paper.
Thus $\sigmav$ in this channel at low velocities can be as large as that at the freeze-out epoch.
Fig.~\ref{fig:sigmav} show that $\sigmav$ for $m_{\chia} > m_t$ basically has a canonical value, $\sim 10^{-26}~\cm^3~\sec^{-1}$.
Nevertheless, the importance of the $t\bar{t}$ channel goes down slowly as $m_{\chia}$ increases, while the importance of the $a_1 h_1$ and $a_1 h_2$ channels slightly goes up.

The $WW$, $ZZ$, $Zh_1$, and $Zh_2$ channels typically appear as minor channels, except for some cases in a mass window of $50~\GeV \lesssim m_{\chia} < m_t$.
This is because $m_{\chia}$ is singlino-dominated and the singlino does not couple to electroweak gauge bosons.
Actually, the $t\bar{t}$ channel is primarily contributed by the $s$-channel $a_1$ process, rather than the $s$-channel $Z$ process.

Searches for high energy muon neutrinos from DM annihilation in the center of the Sun is sensitive to the DM-proton scattering cross section, which is connected to the DM capture process in the Sun and is balanced with the annihilation rate.
For the SD scattering, which is dominant for the capture of $\chia$, the $90\%$ CL exclusion limit from the neutrino telescope IceCube~\cite{Aartsen:2016exj} is stringent than those from direct detection experiments for $m_{\chia} > m_t$, as plotted in Figs.~\ref{fig:SID:c} and \ref{fig:SID:d}.
Note that this limit is derived under the assumption that $\chia\chia$  annihilate into $t\bar{t}$ with a branching fraction of 100\%, which should be a good approximation because $t\bar{t}$ annihilation is dominant for $m_{\chia} > m_t$, as shown in Fig.~\ref{fig:sigmav}. It excludes some points when the $\xi$ factor is not taken into account.

\begin{figure}[!htbp]
\centering
\subfigure[~$m_{\chia}$-$\sigmav$ plane\label{fig:sigmav:a}]{
\includegraphics[width=0.45\textwidth]{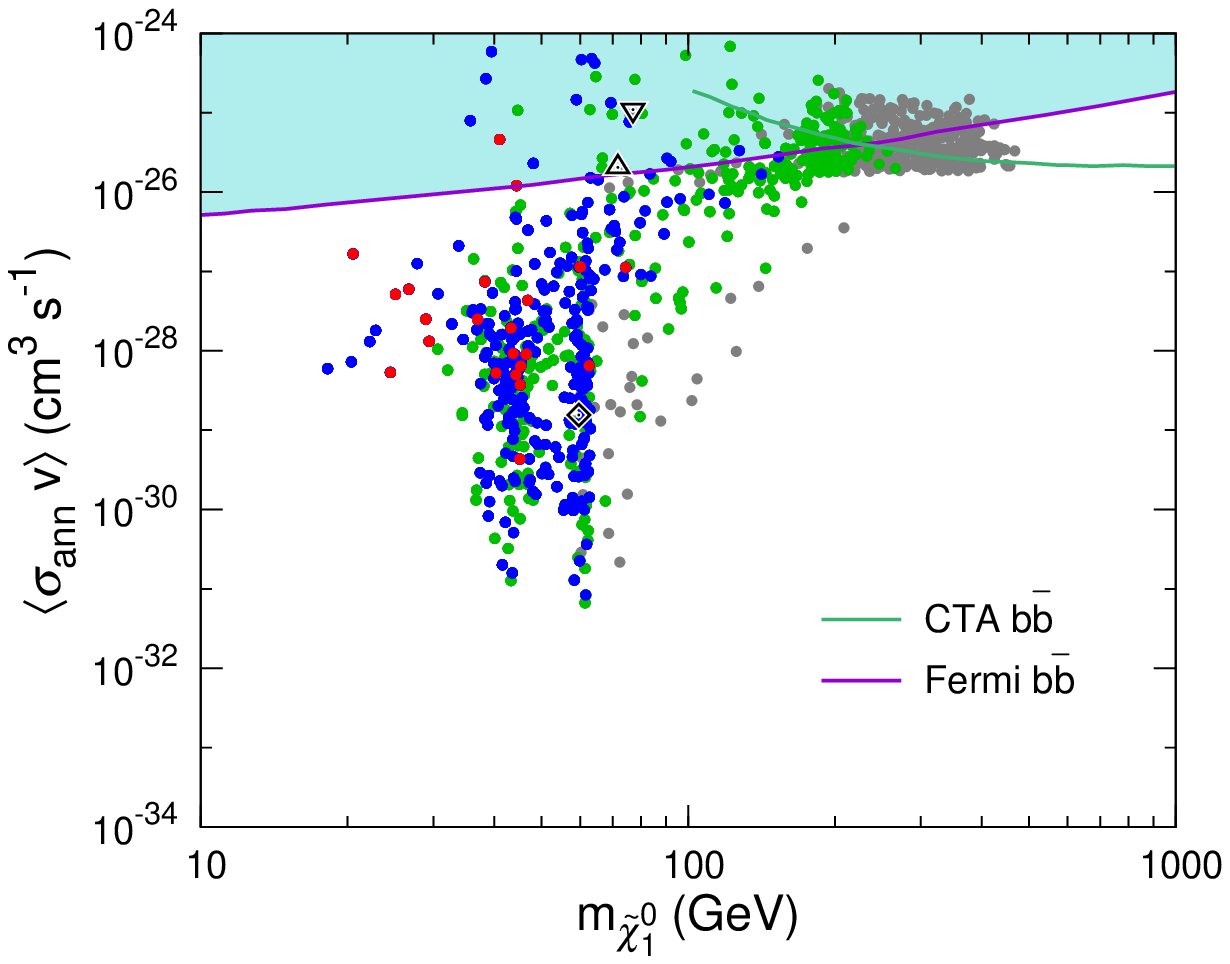}}
\subfigure[~$m_{\chia}$-$\xi^2 \sigmav$ plane\label{fig:sigmav:b}]{
\includegraphics[width=0.45\textwidth]{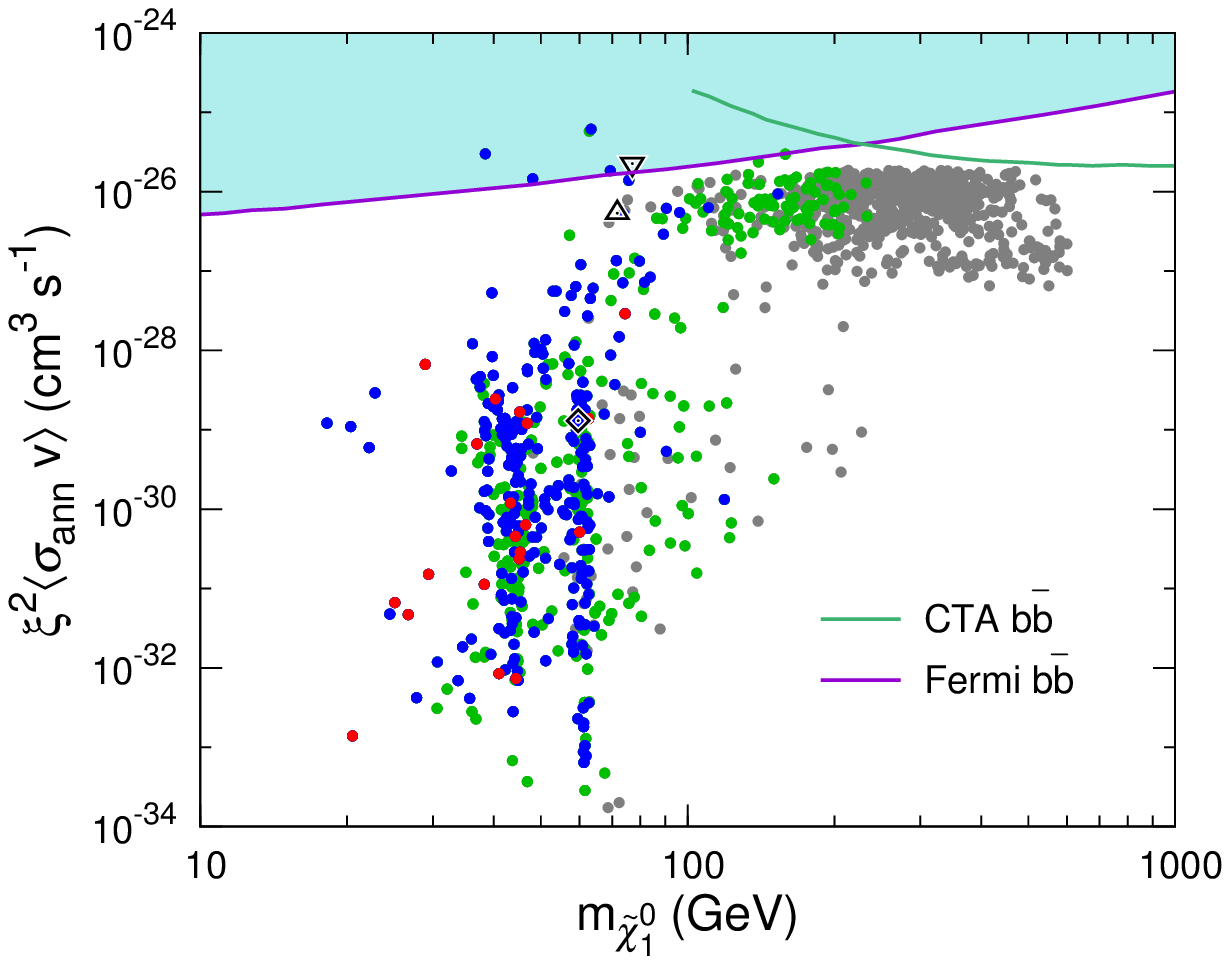}}
\caption{Parameter points projected into the $m_{\chia}$-$\sigmav$ (a)  and $m_{\chia}$-$\xi^2 \sigmav$ (b) planes.
The notation for the points is the same as in Fig.~\ref{fig:SID}.
For a comparison, we also plot the exclusion limit from the Fermi-LAT gamma-ray observation of dwarf galaxies~\cite{Ackermann:2015zua} and the expected exclusion limit for the CTA 100-hour observation of the Galactic Center vicinities~\cite{Doro:2012xx} at 95\% CL, assuming that $\chia$ pairs only annihilate into $b\bar{b}$.
}
\label{fig:sigmav}
\end{figure}

High energy continuous gamma-ray observation is also a robust way to search for nonrelativistic DM signatures. The yield of gamma rays induced by DM depends on the annihilation rate, which is proportional to $\sigmav$ and the square of DM density.
Therefore, when we consider only a fraction $\xi$ of DM is contributed by $\chia$, we should use the reduced annihilation cross section $\xi^2\sigmav$ to compare with experimental results.
Fig.~\ref{fig:sigmav} demonstrates the survived points projected into the $m_{\chia}$-$\sigmav$ and $m_{\chia}$-$\xi^2 \sigmav$ planes with $\sqrt{\left<v^2\right>} = 0.001$.
All annihilation channels are included.
For $m_{\chia} \lesssim 70~\GeV$, because the resonant channels that efficiently worked at the freeze-out epoch are suppressed by the velocity,
$\sigmav$ with $\sqrt{\left<v^2\right>} = 0.001$ has a value of $\sim\mathcal{O}(10^{-31})-\mathcal{O}(10^{-27})~\cm^3~\sec^{-1}$, much smaller than the canonical value.

In Fig.~\ref{fig:sigmav} we also plot the exclusion limit from the Fermi-LAT gamma-ray observation of dwarf galaxies~\cite{Ackermann:2015zua} and the expected exclusion limit for the CTA 100-hour observation of the Galactic Center vicinities~\cite{Doro:2012xx} at 95\% CL.
Both limits are based on the assumption that $\chia\chia$ annihilate into $b\bar{b}$ with a branching fraction of 100\%.
In principle, the gamma-ray spectra induced by different annihilation channels are different.
Here the dominant channels include $b\bar{b}$, $t\bar{t}$, $a_1 h_{1,2}$, $W^+W^-$, and $Z h_{1,2}$.
The gamma-ray spectra from these channels are quite similar~\cite{Cirelli:2010xx}, since they all go through the hadronization process, which is universal, and yield most photons from hadron decays.
Therefore, although the limits are set for the $b\bar{b}$ channel, they should be good approximations for the real situation.
Fig.~\ref{fig:sigmav:a} shows that the Fermi-LAT limit can exclude some points but not so much, while the CTA experiment will be complementary to Fermi-LAT, as it will be more sensitive in the high mass region.
When the $\xi^2$ factor is considered, as shown in Fig.~\ref{fig:sigmav:b}, almost all the points evade these limits.
This means that indirect searches for continuous gamma rays may not be an effective way to explore the singlino-higgsino scenario, compared with direct detection and collider searches.

\section{Conclusions and discussions}
\label{sec:concl}

In this work, we explore the singlino-higgsino scenario in the NMSSM, where the singlino and higgsinos are light and decouple from other superpartners. We assume that the LSP neutralino $\chia$ is singlino-dominated, while $\chib$ and $\chic$ are mainly higgsinos.
Furthermore, the lighter chargino $\chiapm$ is a complete higgsino, with a mass close to $\chib$ and $\chic$.
This setup is distinct from any simplified scenario in the MSSM, as the singlet superfield plays an important role in dark matter phenomenology and collider physics.
In order to satisfy the observed DM relic abundance, the LSP should have either resonant annihilation effects or sizable higgsino components, due to the limited interactions of the singlino.

We carry out a random scan in the parameter space to obtain realistic parameters.
Three benchmark points are picked up to represent typical cases with different neutralino decay modes.
As represented by BP1, in most cases $\chibc$ dominantly decays into $\chia Z$ and $\chiapm$ decays into $\chia W^\pm$.
Therefore, the $3l + \missET$ and $2l + \missET$ searches at the LHC are expected to be sensitive to $\chiapm \chibc$ and $\tilde{\chi}_1^{+} \tilde{\chi}_1^{-}$ direct production with clean SM backgrounds.
We recast the 8~TeV LHC search results and find that the exclusion limit reaches up to $m_{\chib,\chiapm}\sim 250~\GeV$.
Based on a detailed simulation, the prospect of future LHC searches is also investigated. With an integrated luminosity of $30~(300)~\ifb$ at $\sqrt{s}=13~(14)~\TeV$, LHC searches are expected to probe up to $m_{\chib,\chiapm}\sim 320~(480)~\GeV$ and $m_{\chia}\sim 150~(230)~\GeV$.

The $3l + \missET$ and $2l + \missET$ searches lose their sensitivities for the compressed mass spectra where $m_{\chibc,\chiapm} - m_{\chia} \lesssim m_{Z,W}$. This case is typically represented by BP2 and BP3, where $\chibc$ dominantly decays into $\chia h_1$ or $\chia a_1$, while $\chiapm$ decays into off-shell $W$ bosons. Consequently, distinct $3l + \missET$ and $2l + \missET$ final states would not be easily established.
Since the dominant decay channel of $h_1$ and $a_1$ here is $b\bar{b}$, the $2b\text{-jets} + 1l + \missET$ final state provides a particular signature of $\chiapm \chibc$ production.
However, this search channel would be very challenging due to the enormous $t\bar{t}$ background.

Furthermore, we study current bounds and future sensitivities of DM direct and indirect detection experiments.
Unlike collider searches, direct and indirect detection can keep sensitive to heavy LSPs with $m_{\chia} > 250~\GeV$.
Compressed mass spectra are no longer an issue, for instance, BP2 and BP3 have been excluded by current direct detection experiments.
When the LSPs annihilated through a $Z$ or Higgs resonance in the early Universe to achieve an acceptable relic abundance, its higgsino components could be very tiny, leading to small DM-nuclei scattering cross sections as well as small nonrelativistic annihilation cross sections.
Thus, this is a difficult case for direct and indirect searches.
Fortunately, most parameter points in this case would be covered by the LHC $3l + \missET$ and $2l + \missET$ searches, as long as $m_{\chib,\chiapm}\lesssim 480~\GeV$.
Therefore, we conclude that the singlino-higgsino scenario will be very well investigated in near future LHC searches and DM detection experiments.

\begin{acknowledgments}
This work is supported by the National Natural Science Foundation of China under Grant Nos.~11475189,~11475191,~11135009, the 973 Program of China under
Grant No.~2013CB837000, and
by the Strategic Priority Research Program
``The Emergence of Cosmological Structures'' of the Chinese
Academy of Sciences under Grant No.~XDB09000000.
ZHY is supported by the Australian Research Council.
\end{acknowledgments}


\bibliographystyle{JHEP}
\bibliography{NMSSM}
\end{document}